\documentclass[a4paper, 11pt]{article}

\usepackage{amsmath, amsfonts, amssymb, amsthm}
\usepackage{mathtools}
\usepackage{commath}               
\usepackage{paralist}
\usepackage{graphicx}  
\usepackage[T1]{fontenc}
\usepackage{float}

\title{Deep Neural-network Prior for Orbit Recovery from Method of Moments}
\author{Yuehaw Khoo$^1$ \and Sounak Paul$^1$\footnote{Corresponding author} \and Nir Sharon$^2$}

\date{
	\small{$^1$Department of Statistics, University of Chicago, Chicago, USA \\ \texttt{\{ykhoo, paulsounak96\}@uchicago.edu}\\
	$^2$School of Mathematical Sciences, Tel Aviv University, Tel Aviv, Israel} \texttt{nir.sharon@math.tau.ac.il}\\[2ex]
}

\usepackage{multirow}
\usepackage{algorithmic}
\usepackage{algorithm}
\usepackage{float}
\usepackage{wrapfig}
\usepackage{mdframed}
\usepackage{xcolor}
\usepackage{bm}
\usepackage[title]{appendix}
\usepackage{resizegather}
\usepackage{bbm}
\usepackage{subcaption} 
\usepackage{wrapfig}
\usepackage{epsfig}     
\usepackage{siunitx}
\usepackage{nicefrac}
\usepackage{xurl}
\usepackage{enumitem}
\usepackage{geometry}
\usepackage{tikz-cd} 
\usepackage{hyperref}  
\hypersetup{colorlinks=true, urlcolor=black, citecolor=red}
\hypersetup{pdfauthor={Name}}  

\newcommand{\so}[1]{{\text{SO}\negmedspace\left(#1\right)}}

\newcommand{\ep}{\varepsilon}

\newcommand{\angstrom}{\text{\normalfont\AA}}

\theoremstyle{definition}

\newcommand\rev[1]{\textcolor{black}{#1}}

\begin{document}

    \maketitle
    
    \begin{abstract}
        Orbit recovery problems are a class of problems that often arise in practice and various forms. In these problems, we aim to estimate an unknown function after being distorted by a group action and observed via a known operator. Typically, the observations are contaminated with a non-trivial level of noise. 
        
        Two particular orbit recovery problems of interest in this paper are multireference alignment and single-particle cryo-EM modeling. In order to suppress the noise, we suggest using the method of moments approach for both problems while introducing deep neural network priors. In particular, our neural networks should output the signals and the distribution of group elements, with moments being the input. In the multireference alignment case, we demonstrate the advantage of using the NN to accelerate the convergence for the reconstruction of signals from the moments. Finally, we use our method to reconstruct simulated and biological volumes in the cryo-EM setting.
        
    \end{abstract}
    



 

	

\section{Introduction}
Orbit recovery refers to a type of estimation problems that involve incorporating the effect of a group on a data model. The resulting solution is determined up to an arbitrary group action, meaning that the solution forms an orbit. This class of estimation problems is crucial in various fields of science and engineering, ranging from signal processing to structural biology. For instance, medical tomography often collects imaging data that undergoes unknown transformations. Along with pixel-wise noise, each image may experience rotation, translation, flipping, or other group actions in an unknown manner. This article examines two problems in this category and proposes a new approach to solving them.

The first issue we discuss is Multi-Reference Alignment (MRA), which involves estimating a signal from the observation of noisy, circularly shifted copies of it. This model, which has its origins in both signal processing~\cite{zwart2003fast} and structural biology~\cite{scheres2005maximum, theobald2012optimal}, provides a foundation for exploring the relationship between the group structure, noise levels, and the possibility of recovery ~\cite{wein2018statistical, abbe2018multireference, moitra2019spectral}. The second problem we consider is 3D volume reconstruction in Cryogenic Electron Microscopy (cryo-EM), as discussed in~\cite{sharon2020method}. In cryo-EM, the goal is to retrieve a 3D volume from 2D noisy images that result from rotating the volume and applying a fixed tomographic projection. The outcome is a set of 2D images, which are usually heavily contaminated with noise.

The Method of Moments (MoM) is a classical estimation technique that has been adapted in modern forms to provide a powerful computational tool for solving large-scale problems, especially when dealing with high noise levels. The MoM consists of two stages. First, we compute the observable moments by averaging the low-order statistics of any observation.
The second stage involves retrieving the required signal from the observable moments by analyzing the relationship between the observable and analytical moments, applying moments-fitting, and deriving the unknown parameters from it. This second stage is the focus of this study.
\rev{The usage of MoM is advantageous in several ways. Its robustness is derived from the fact that noise is averaged out during the computation of observable moments. Namely, given enough data, the effect of noise can be rendered insignificant. Also, MoM gleans information about the data only through the moments, so it does not require multiple passes over the data set. This is beneficial while dealing with huge data sets, as the moment calculation from the data takes place only in the first stage and in one pass~\cite{sharon2020method, singerICM2018}. However, this method does have a major drawback. We can lose resolution since we are not using information from all the moments. MoM thus leads to low-dimensional reconstructions. Fortunately, our focus is mainly to recover an \textit{ab-initio} model; hence a low-dimensional reconstruction suffices. In the case of cryo-EM, this \textit{ab-initio} model is used as an initialization for iterative refinement algorithms, where reconstruction enables several possible conformations by further refinement~\cite{frank2006three}.}

This paper introduces a new version of the MoM that incorporates a neural-network prior to tackling orbit recovery problems. In particular, we demonstrate the effectiveness of the amortized MoM for the two orbit recovery problems discussed earlier: Multi-Reference Alignment (MRA) and single-particle Cryogenic Electron Microscopy (cryo-EM) modeling. Learning algorithms have recently taken a central role in cryo-EM computational methods: a deep neural network for modeling continuous heterogeneity (3DFlex)~\cite{punjani2021advances}, ab initio neural reconstruction~\cite{zhong2021cryodrgn, zhong2021cryodrgn2}, and many other parts of the cryo-EM pipeline~\cite{jimenez2021deepalign, bepler2020topaz, kimanius2021new}, to name a few. Moreover, amortized learning has recently appeared in a study for 3D modeling in cryo-EM~\cite{levy2022cryoai}. However, noise resilience remains one of the most significant challenges in cryo-EM 3D reconstruction. The proposed ``amortized'' MoM technique provides a genuine alternative that addresses this challenge effectively while also addressing the additional challenge of ever-growing cryo-EM datasets.

In our method, we treat the group elements of each problem as random variables and consider them as nuisance parameters or latent variables. Rather than estimating them directly, we aim to target their density function along with the unknown signal. Our MoM incorporates neural networks to approximate the signal and distribution to achieve this. We demonstrate that in the case of MRA, a neural network can encode existing algorithms for solving the inverse problem from the moments. Moreover, we propose that the moment inversion process can be significantly improved by initializing the neural networks with those trained in a supervised manner on similar instances of the recovery problem. Our approach to the MRA problem serves as a proof-of-concept, and we extend these techniques to the case of cryo-EM.

The paper is organized as follows. Section~\ref{sec:notation} describes the problem formulation. Then, in Section~\ref{prob_formulate}, we present the method of moments approach for the MRA model and cryo-EM model individually as special cases of our class of estimation problems~\eqref{eqn:general_formulation}. Next, Section~\ref{nn_prior} introduces neural network priors for representing the volume and distribution of group elements for both models. Next, Section~\ref {numeric_eg} illustrates the performance of our neural network priors in the reconstruction of various simulated as well as real-world biological volumes.
Finally, we conclude with Section~\ref{final_thoughts}, including a summary of the next steps in this exciting line of research. 

\section{Problem formulation: orbit recovery} \label{sec:notation}

Let $v$ be an unknown scalar-valued object defined as a function
\begin{equation} \label{eqn:the_signal}
v \colon  \Omega \to \mathbb{R} , 
\end{equation}
and let $G$ be a group with a well-defined action on $v$, that is $G \curvearrowright \Omega$. One class of estimation problems we are concerned with consists of the following general formulation. Our goal is to estimate the function $v$, where we observe $N$ samples,
\begin{equation} \label{eqn:general_formulation}
v_j  = \mathcal{A}(g_j \circ v) + \varepsilon_j,  \quad g_j \sim \rho, \quad j=1,\ldots, N , 
\end{equation}
where  $\left\lbrace \varepsilon_j \right\rbrace_{j=1}^N$ is a set of i.i.d.\ random noise terms, $\mathcal{A}$ is a known operator, and  $\{ g_j \}_{j=1}^N$ is a set of i.i.d. random group elements distributed according to some distribution $\rho$ on $G$. These are treated as latent variables or nuisance parameters for our problem since the objective is to get $v$. Note that one can only estimate $v$ up to a group action, since for any estimator $\hat v$ and  $\{ \hat{g}_j \}_{j=1}^N$ for the object and latent group elements respectively, $g\circ \hat v$ and $\{ \hat{g}_j g^{-1} \}_{j=1}^N$ give another set of equivalent estimators for any fixed $g\in G$. Hence our goal becomes the \textit{orbit recovery} of $v$.

Customarily, in the lower-noise regime, where the magnitude of $\ep_j$ is smaller than the magnitude of $v$, a solution to~\eqref{eqn:general_formulation} is obtained using the following scheme. First $g_{ij} \approx g_i g_j^{-1}$ is estimated from $v_i$ and $v_j$. Then one recovers the group elements $ \{ g_j \}_{j=1}^N$ from the set of their ratios $\{ g_i g_j^{-1} \}_{i,j=1}^N$, i.e. solving a \textit{synchronization problem over $G$}. Then with a good estimation for $ \{ g_j \}_{j=1}^N$, we solve for $v$ in problem~\eqref{eqn:general_formulation} via solving a linear system of equations~\cite{anden2018structural}. 

As the level of noise in the observations increases, the random noise heavily influences the alignment results so that even with the ground truth $v$ given, one would often be fooled to assign wrong group actions with large errors~\cite{shkolnisky2012viewing, singerICM2018}. A different approach consists of treating the group elements $\{g_{j}\}_{j=1}^N$ as nuisance parameters and having the signal be the primary estimation target. Therefore, when considering a high level of noise, we focus on methods that marginalize over the nuisance parameters by treating them as random variables~\cite{berger1999integrated}. The estimation of $v$ can be done via maximizing the marginalized posterior distribution that has $v$ being the random variable or using a method of moments with moments formed by averaging $v_j$'s such that there is no dependency on $g_j$'s.

\section{Method of moments}\label{prob_formulate}

The method of moments (MoM) is a classical technique to estimate parameters from observed statistics. Two recent models where MoM was already successfully employed are multireference alignment (MRA)~\cite{abbe2018multireference, bendory2017bispectrum} and cryo-EM recovery~\cite{sharon2020method}, where the operator $\mathcal{A}$ of~\eqref{eqn:general_formulation} is the identity and a tomographic projection, respectively. The group consists of circular shifts on MRA and 3D rotations in cryo-EM recovery. Then, the $m$-th moment is the expectation of the $m$-th-order tensor product of the samples with themselves, i.e., $v_j^{\otimes m }$. Interestingly, the minimal number of moments to guarantee uniqueness also determines the sample complexity --- the number of samples needed, as a function of noise level, in order to have a consistent estimation, see ~\cite{abbe2018multireference, abbe2018estimation, perry2019sample}. Therefore, when studying~\eqref{eqn:general_formulation}, the MoM plays a significant role as a baseline for designing computational algorithms and analyzing the sample complexity.

\subsection{The MRA model}\label{sec: MRA model}

We begin with the MRA model,~\eqref {eqn:general_formulation}, where $\mathcal{A}$ is the identity. In this situation, the unknown signal $v$ is defined on a unit, symmetric segment $\mathcal{I}=[-\frac{1}{2}, \frac{1}{2}]$. Namely, the signal is $v: \mathcal{I} \rightarrow \mathbb{R}$, and we further assume it is a periodic, band-limited function. Let $G$ be the group of circular translations (rotations) on $\mathcal{I}$, whose elements $s_j$ shift $v$ in the following manner,
\begin{equation}
    s_j \circ v := v(\cdot - s_j).
\end{equation}
Here, we interpret the difference as modulo the segment, namely $\cdot - s_j$ is always in $\mathcal{I}$.

We next formulate the MRA problem in the Fourier domain. For convenience, we discuss the case when there is no noise. Let $\widehat v_j$ be the Fourier transform of $v_j$, in this case, a shift $s_j$ becomes a phase, i.e.
\begin{equation}\label{eq:translated signal MRA}
    \widehat v_j(k) =  \exp(iks_j) \widehat v(k), \ k\in[-\pi,\pi].
\end{equation}
The frequency $k$ has a natural bandlimit $\vert k \vert \leq \pi$ since the signal $v_j$ is usually provided on $n$ discretized points in $\mathcal{I}$, where $n$ is chosen to satisfy its Nyquist frequency. As for our observation, let $K_{1}$ be the set of $n$ equispaced points between $[-\pi,\pi]$. Then, we have 
\begin{equation}\label{mra_nonoise}
    \widehat v_j(k)  = \exp(ik s_j) \widehat v(k), \quad k \in K_{1}.
\end{equation}
Henceforth, for brevity, we use $\widehat v_j(K_1)  = \exp(iK_{1} s_j)\odot \widehat v(K_1)$ instead of the pointwise notation, where ``$\odot$'' denotes the Hadamard product.

Finally, in MoM for MRA, we let 
\begin{equation}\label{eqn:MRA moments def}
    M_F^1[\widehat v,\rho](k_1) = \mathbb{E}_\rho\left(\frac{1}{N}\sum^N_{j=1} \widehat v_j(k_1)\right), \quad 
    M_F^2[\widehat v,\rho](k_1, k_2) = \mathbb{E}_\rho\left(\frac{1}{N}\sum^N_{j=1} \widehat v_j(k_1) \widehat v_j(k_2)^*\right).
\end{equation}
Here, $M_F^1$ and $M_F^2$ are functions of $\hat v,\rho$. The goal is to retrieve $\widehat v$ from unbiased estimators $\hat M_F^1,\hat M_F^2$ of $M_F^1, M_F^2$ when having noisy data via matching the moments. 

\subsection{The cryo-EM model}

The problem~\eqref{eqn:general_formulation} also serves as a simplified model of single-particle cryo-EM, where the operator $\mathcal{A}$ is a tomographic projection along a fixed axis. Cryo-EM is a prominent method for determining the high-resolution 3-D structure of biological macromolecules from its 2-D noisy projection images~\cite{frank2006three}.

For the cryo-EM model, we denote by $v \colon \mathbb{R}^3 \to \mathbb{R}$ the Coulomb potential of the 3D volume we aim to determine, where we assume that $v$ is compactly supported in a ball of radius $\frac{1}{2}$ around the origin, that is inside $\mathcal{I}^3$. We define the composition of $R_j$ with the volume $v$ as
\begin{equation} \label{eqn:rotation}
R_j\circ v\ \left(x,y,z\right) \ =\ v \left(R_j^T [x\ y\ z]^T\right), \quad (x,y,z)\in \mathcal{I}^3,
\end{equation}
viewing $R_j$ as a $3\times 3$ matrix in the right hand side of~\eqref{eqn:rotation} since $\so{3}\subset \mathbb{R}^{3\times 3}$. Let $\mathcal{P} \colon \mathbb{R}^3 \to \mathbb{R}^2$ be the operator that projects a 3D volume along the $z$ axis to a 2D image, i.e.
\begin{equation} \label{eqn:projection}
\mathcal{P} \circ v\ \left(x,y\right) \ =\ \int_{-\infty}^{\infty} v \left(x, y, z\right) dz, \quad (x,y,z)\in \mathcal{I}^2.
\end{equation}
Then, a standard image formation model in the absence of noise, after filtering the effect of the contrast transfer function (CTF), image cropping, and centering, is (see~\cite{frank2006three, heimowitz2021centering}), 
\begin{equation}\label{eqn:cryoem_case}
v_j = \mathcal{P}\circ R_j \circ v, \quad j=1,\ldots,N,
\end{equation}
where $R_j \in \so{3}$ are the unknown group elements. To avoid the computationally intensive integration in~\eqref{eqn:projection}, we reformulate our problem in the Fourier domain. There, we can exploit the Fourier Slice Theorem to speed up computation significantly. We define $\widehat v \colon [-\pi,\pi]^3 \rightarrow \mathbb{C}$ as the Fourier transform of $v$, and $S\colon [-\pi,\pi]^2 \rightarrow \mathbb{C}$ as the slice operator given as
\begin{equation} \label{eqn:slice_operator}
    S \circ \widehat v (k_x, k_y)\ =\ \widehat v \left(k_x, k_y, 0\right),
\end{equation}
i.e., $S \circ \widehat v$ is obtained by slicing $\widehat v$ across the plane given by $z=0$. Then, the Fourier Slice Theorem states that:
\begin{equation} \label{eqn:fourier_slice}
    \mathcal{F}_{2D} \circ \mathcal{P} \circ R\ =\ \mathcal{S} \circ R \circ \mathcal{F}_{3D},
\end{equation}
where $R\in \so{3}$, $\mathcal{F}_{2D}$ and $\mathcal{F}_{3D}$ are the 2D and 3D Fourier transformations, respectively. Therefore, in the no-noise setting, the equivalent of~\eqref{eqn:cryoem_case} becomes,
\begin{equation} \label{eqn:cryo_nonoise}
    \widehat v_j(k_x, k_y)\ =\ S\circ R_j\circ \widehat v(k_x, k_y),\quad (k_x, k_y)\in [-\pi,\pi]^2,
\end{equation}
where $\widehat v_j$ is the Fourier transform of $v_j$. Let $K_2$ be a grid of $n^2$ equispaced points on $[-\pi,\pi]^2$, flattened as a one-dimensional vector. Now our observations are
\begin{equation} \label{eqn:cryo_on_grid}
    \widehat v_j(K_2)\ =\ S\circ R_j\circ \widehat v(K_2),
\end{equation}
and the associated moments are
\begin{equation} \label{cryo_moments}
\begin{split}
    M_F^1[\widehat v,\rho](k_x,k_y) &= \mathbb{E}_\rho\left(\frac{1}{N}\sum^N_{j=1} \widehat v_j(k_x, k_y)\right)\\
    M_F^2[\widehat v,\rho](k_x,k_y,k'_x,k'_y) &= \mathbb{E}_\rho\left(\frac{1}{N}\sum^N_{j=1} \widehat v_j(k_x, k_y) \widehat v_j(k'_x, k'_y)^* \right).
\end{split}
\end{equation}
We aim to retrieve $\widehat v$ by matching the moments $M_F^1[\hat v,\rho](K_2,K_2)$ and $ M_F^2[\hat v,\rho](K_2,K_2)$ with some unbiased estimators $\hat M_F^1,\hat M_F^2$ when having noisy data.

\section{Neural network priors for method of moments }\label{nn_prior}

This section presents neural network (NN) approaches for reconstructing the signal $v$ and distribution $\rho$ in MRA and cryo-EM settings. The general strategy is to view both the signal and distribution as being mapped by a NN from the estimated moments $\hat M_F^1$ and $\hat M_F^2$, as various previous works have shown that \rev{for MRA, $\hat M_F^1,\hat M_F^2$ are generically sufficient statistics for estimating the unknowns~\cite{abbe2018multireference}, while for cryo-EM, they have enough information for recovering a low-resolution reconstruction~\cite{sharon2020method}.}
In the MRA case, we design an encoder that can map the empirical moments to discretized signal and density. In the cryo-EM case, we further design an encoder-decoder structure that allows us to take the moments as input and give a continuous representation of a 3D volume.

\subsection{NN for MRA}\label{amortized_mra}
\begin{figure}
    \includegraphics[width = 15cm]{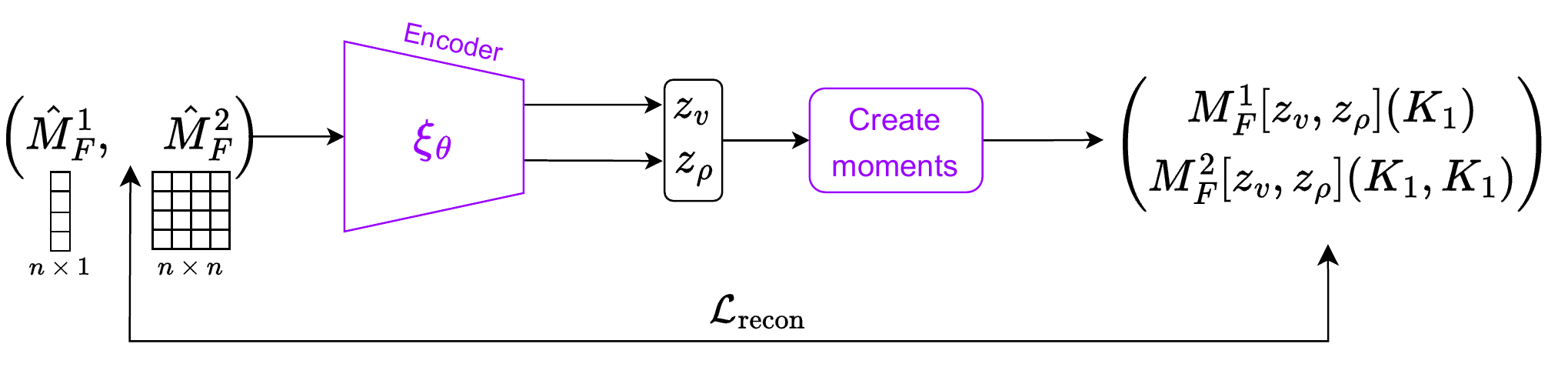}
    \caption{\textbf{Overview of our MRA pipeline}: The encoder $\xi_\theta$ takes moments $(\hat M_F^1,\hat M_F^2)$ as input, and outputs $z_\rho\in \mathbb{R}^n$, approximating a discretized probability density $\rho(X_1)$, and $z_v\in \mathbb{R}^n$ that approximates a discretized Fourier signal $\widehat v(K_1)$. Next, we use $z_\rho$ and $z_v$ to create $\left(M_F^1[z_v, z_\rho](K_1), M_F^2[z_v, z_\rho](K_1, K_1)\right)$ via equation~\eqref{eqn:recon moments}, which we then compare with the inputs to the encoder, i.e., $(\hat M^1,\hat M^2)$ via the loss function $\mathcal{L}_\textnormal{recon}$~\eqref{eqn:mra_loss}.}
    \label{fig:mra_diag}
\end{figure}

In the MRA case, we want to train a NN that can take the moments as inputs and give the signal and density as outputs, which can further be used to initialize an iterative reconstruction algorithm. More precisely, we define $F\in \mathbb{C}^{n\times n}$ as the matrix representation of a normalized Fourier transform where $F^* F = I_n$, and $X_1, K_1$ as sets of $n$ equispaced points on $\mathcal{I} = \left[-\frac{1}{2}, \frac{1}{2}\right]$ and $[-\pi,\pi]$ respectively. The main component is an encoder, i.e., a neural network $\xi_\theta$, that takes the moments $\hat M_F^1, \hat M_F^2$ as inputs and outputs $z_\rho, z_v\in \mathbb{R}^n$, where $z_\rho$ approximates a discretized density $\rho(X_1)$ and $z_v$ approximates a discretized signal $\widehat v(K_1)$.

The encoder $\xi_\theta:=(\xi_\theta^v,\xi_\theta^\rho)$ consists of two NN $\xi_\theta^v$ and $\xi_\theta^\rho$, which are two 1D convolutional NNs (CNNs) that take $\hat M_F^1\in \mathbb{C}^{n}, \hat M_F^2\in \mathbb{C}^{n\times n}$ as input vector fields supported on $n$ grid points. Figure \ref{fig:mra_diag} provides an overview of our pipeline for MRA. While the details of the architectures are provided in \ref{architecture}, here we provide motivations as for how a CNN has the capability to learn a mapping from the moments $\hat M_F^1\in \mathbb{C}^{n}, \hat M_F^2\in \mathbb{C}^{n\times n}$ to $\widehat v(K_1)$. For simplicity, suppose $\vert \widehat v(k) \vert = 1$. Using the definitions in \eqref{eqn:MRA moments def} and the fact that translating $v$ by $s$ is equivalent to letting $\widehat v(k) \rightarrow \widehat v(k) \exp(iks)$, one can show that
\begin{equation}
    M_F^2[\widehat v,\rho](k_1,k_2) = \widehat v(k_1) \widehat \rho(k_1-k_2)\widehat v(k_2)^*,
\end{equation}
as in~\cite{abbe2018multireference}. In this case, $M_F^2[\widehat v,\rho](K_1,K_1)$ admits the eigendecomposition
\begin{eqnarray}\label{eqn:circulant M2}
M_F^2[\widehat v,\rho](K_1,K_1) &=&  \text{diag}(\widehat v(K_1)) F^* ( F [\widehat \rho(k_1-k_2)]_{k_1,k_2}F^*) F \text{diag}(\widehat v(K_1)^*)\cr 
&=&  [\text{diag}(\widehat v(K_1)) ]F^* \text{diag}(\rho(X_1)) [F \text{diag}(\widehat v(K_1)^*)]
\end{eqnarray}
since $[F \text{diag}(\widehat v(K_1)^*)]$ is an orthogonal matrix (due to the assumption $\vert \widehat v(k) \vert = 1$). From this form, it is clear that the eigenvalues of $M_F^2[\hat v,\rho](K_1,K_1)$ are $\rho(X_1)$ and furthermore, the eigenvectors are $F \text{diag}(\widehat v(K_1)^*)$. Since the spectral information of the second moments contains information concerning the signal and density, if an NN can mimic a spectral method, then it can learn the mapping from moments to the signal and density. 

The form of $M_F^2[\widehat v,\rho](K_1,K_1)$ in \eqref{eqn:circulant M2} suggests that it is a circulant matrix. Therefore if we want to devise a neural network that takes $\hat M_F^2 = M_F^2[\widehat v,\rho](K_1,K_1)$ (when there is no noise) as input and output the eigenvectors $F \text{diag}(\widehat v(K_1)^*)$, we can have a neural network, composed of 1D convolutional layers, that takes $\hat M_F^2$ as a 1D $n$-dimensional vector field supported on $n$ grid points. For example, to get an eigenvector of $M_F^2[\widehat v,\rho](K_1,K_1)$, a convolutional layer $l_1:\mathbb{C}^n\rightarrow \mathbb{C}^n$ can take the form
\begin{equation}\label{eqn:conv}
l_1(u) = \frac{\hat M_F^2 u}{\|\hat M_F^2 u\|_2}.
\end{equation}
One can think about $\hat M_F^2$ as the weights of the convolutional layer $l_1$, and the division by $\|\hat M_F^2 u\|_2$ as some nonlinearities in the NN. Repeated applications of $l_1$, gives an eigenvector of  $M_F^2[\widehat v,\rho](K_1,K_1)$. After obtaining an eigenvector, say $F(:,1)\widehat v(K_1(1))$ where $F(:,1)$ is the first column of $F$, the NN can simply apply some pointwise nonlinearities layer $l_2:\mathbb{C}^n\rightarrow \mathbb{C}^n$ that performs
\begin{equation}
    l_2(u(i)) = \frac{u(i)}{F(i,1)}, \quad i\in [n].
\end{equation}
 Putting these elements together into a deep NN, i.e., $l_2\circ l_1\circ\cdots\circ l_1$ should give $\widehat v(K_1(1))$. Similar operations can be carried out for other eigenvectors. We also use a similar structure for $\xi_\theta^\rho$ to output $z_\rho$ that approximates $\rho(X_1)$, since it is clear that if $u = l_2\circ l_1\circ\cdots\circ l_1(\hat M_F^2)$ is an eigenvector of $\hat M_F^2$, applying another nonlinearity 
 \begin{equation}
     l_3(u) = \langle u, \hat M_F^2 u\rangle
 \end{equation}
gives the eigenvalue of $\hat M_F^2$ which contains information of $\rho(X_1)$ (as shown in \eqref{eqn:circulant M2}).

\subsection{NN for cryo-EM}\label{sec:amortized_cryo}

\begin{figure}
    \includegraphics[width = 15cm]{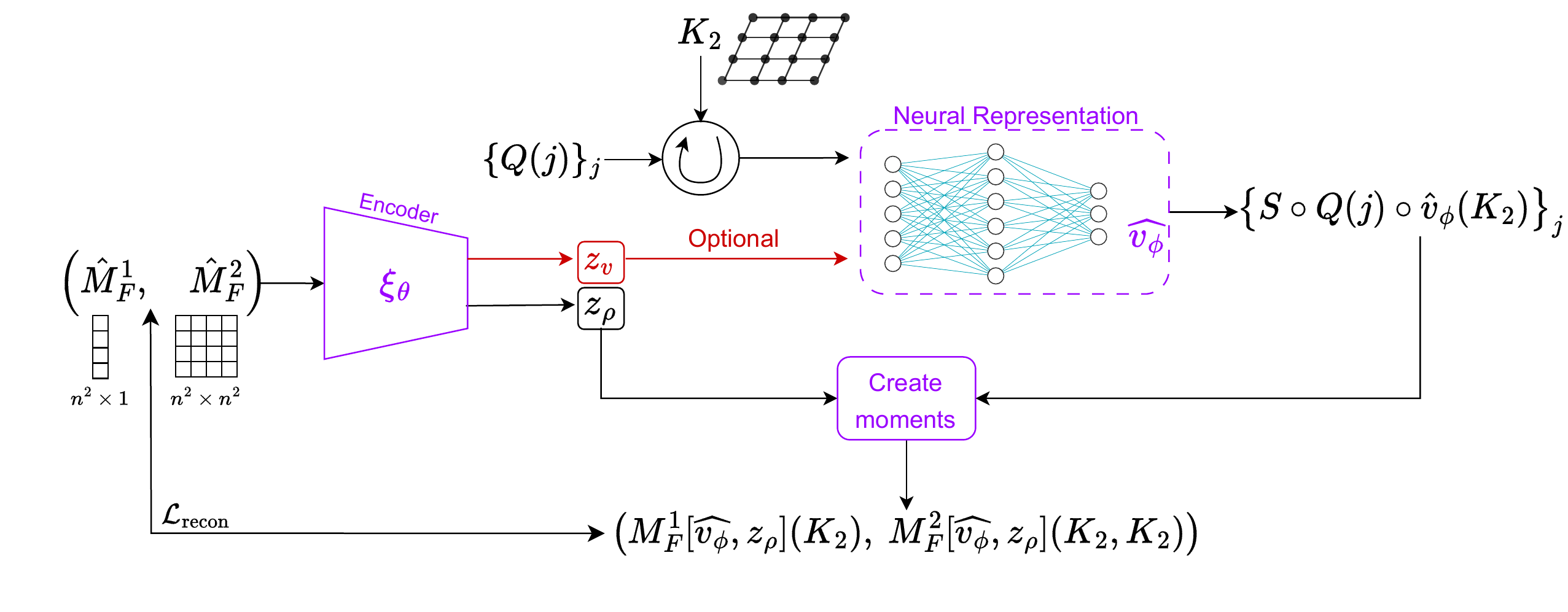}
    \caption{\textbf{Overview of our cryo-EM pipeline}: The encoder $\xi_\theta$ takes moments $(\hat M_F^1,\hat M_F^2)$ as input, and outputs $z_\rho\in \mathbb{R}^{|Q|}$, approximating a discretized probability density $\left(\rho(R) \right)_{R \in Q}$ for some fixed set of quadrature points $Q\subset\so{3}$. Next, we create copies of the grid $K_2$ rotated corresponding to the elements of $Q$ and input them to our neural representation $\widehat v_\phi$, which outputs corresponding slices of a running estimate of $\widehat v$. These slices $\left\{S\circ Q(j)\circ \widehat v_\phi (K_2)\right\}_j$ along with $z_\rho$ are used to create $\left(M_F^1[\widehat v_\rho, z_\rho](K_2), M_F^2[\widehat v_\rho, z_\rho](K_2, K_2)\right)$ via equation~\eqref{eqn:quadrature_moments}, which we then compare with the inputs to the encoder, i.e., $(\hat M_F^1,\hat M_F^2)$ via the loss function $\mathcal{L}_\textnormal{recon}$ in~\eqref{eqn:cryo_loss}. Optionally, $\xi_\theta$ can also be used to output an extra $z_v$, a latent variable of $\widehat v$ that can be inputted into $\widehat v_\phi$.}
    \label{fig:cryoem_diag}
\end{figure}

We make some alterations to our MRA architecture for cryo-EM reconstruction since we need to output a continuous representation of the volume to facilitate computing the moments involving the reconstructed volume. Just as in the case of MRA, we have an encoder $\xi_\theta^\rho$ that outputs information regarding the density. More precisely, let $Q\subset \so{3}$ be a set of quadrature points on $\so{3}$ and $q=\vert Q \vert$. We want $\xi_\theta^\rho:(\hat M_F^1, \hat M_F^2)\rightarrow z_\rho$ $(\hat M_F^1, \hat M_F^2)$ are estimators of \eqref{cryo_moments}) where $z_\rho$ should approximate $\left(\rho(R) \right)_{R \in Q}$ and $\rho$ is a density on $\so{3}$. 

However, unlike the case of MRA, we now want to have a continuous representation of the Fourier volume where the benefit is explained as follows. Let $\widehat v_\phi:\mathbb{R}^3\rightarrow \mathbb{C}$ be an NN that represents a volume on the Fourier domain, and $K_2$ be $n^2$ equispaced points on $[-\pi,\pi]^2$. Suppose $\widehat v_\phi = \widehat v$ and $z_\rho = \rho(Q)$, one can evaluate $M_F^1[\widehat v, \rho](K_2,K_2), M_F^2[\widehat v, \rho](K_2,K_2)$ defined in \eqref{cryo_moments} approximately via the quadrature rule
\begin{equation} \label{eqn:quadrature_moments}
    \begin{split}
         M_F^1[\widehat v_\phi, z_\rho](K_2) & \approx \sum_{j=1}^q z_\rho(j) S\circ Q(j)\circ \widehat v_\phi(K_2) ,\\
    M_F^2[\widehat v_\phi, z_\rho](K_2, K_2) &\approx \sum_{j=1}^q z_\rho(j) \left(S\circ Q(j)\circ \widehat v_\phi(K_2)\right) \otimes \left(S\circ Q(j)\circ \widehat v_\phi(K_2)\right),
    \end{split}
\end{equation}
where, by an abuse of notation, we think about $z_\rho=\rho(Q)$, i.e., the density $\rho$ discretized on $Q$, as $\rho$ itself and $Q(j)$ is an element in the set $Q$. For simplicity, in this paper, we consider a quadrature rule with uniform quadrature weights, as seen in~\eqref{eqn:quadrature_moments}. It is clear that having a continuous $\widehat v_\phi$ allows us to obtain $\widehat v_\phi(Q(j)^T (k_x,k_y,0))$ for any $(k_x,k_y)\in K_2$ easily. 

Note that we also allow the flexibility to have an encoder $\xi^v_\theta$ just as in the case of MRA. In this case, $\xi^v_\theta: (\hat M_F^1, \hat M_F^2)\rightarrow z_v$ where $z_v$ is some latent variable of the volume. In this case, we simply let $\widehat v_\phi: \mathbb{R}^{3+\vert z_v\vert}  \rightarrow \mathbb{C}$ where the extra inputs of $\widehat v_\phi$ corresponds to the output of $\xi^v_\theta$. The neural network pipeline we devise is shown in Figure~\ref{fig:cryoem_diag}, where $\xi_\theta = \left(\xi_\theta^\rho, \xi_\theta^v\right)$. As for the architecture of $\xi^v_\theta,\xi^\rho_\theta$, we adopt the type of architecture we use in Section~\ref{amortized_mra}, though one should be able to improve it according to the structure of the cryo-EM problem. The details of $\xi_\theta$ and $\widehat v_\phi$ are given in \ref{architecture}.

\section{Numerical examples}\label{numeric_eg}

This section presents the results of numerical experiments with our Method of Moments algorithm with NN prior done using PyTorch~\cite{NEURIPS2019_9015}.

\subsection{MRA}

We first present results using the method for MRA in Section~\ref{amortized_mra}. There are two phases when using the NN detailed in Section~\ref{amortized_mra}: training phase consisting of supervised learning (Section~\ref{section: training phase MRA}) and reconstruction phase consisting of unsupervised learning (Section~\ref{section: recon phase MRA}). 

For evaluation purposes, we define the reconstruction error (also referred to as relative error) of an estimator $u\in\mathbb{R}^n$ of a signal $v$ (or a distribution $\rho$) discretized at $X_1$, to be
\begin{equation}\label{eq:relative error v mra}
    \inf_{s\in \mathcal{I}} \frac{\norm{s\circ v\ \left(X_1\right)\ -\ u}_F}{\norm{v \left(X_1\right)}_F},\quad \inf_{s\in \mathcal{I}} \frac{\norm{s\circ \rho\ \left(X_1\right)\ -\ u}_F}{\norm{\rho \left(X_1\right)}_F}.
\end{equation}
Note that in this case, we identify the group with $X_1$. In addition, we define the relative errors for any moment estimators $A_1, A_2$ for the first and second moments, respectively, as 
\begin{equation}\label{eq:relative error moments}
\frac{\norm{M_F^1[\widehat v,\rho](K_1) - A_1}_F}{\norm{A_1}_F}, \quad\frac{\norm{M_F^2[\widehat v,\rho](K_1,K_1) - A_2}_F}{\norm{A_2}_F} .
\end{equation}

\subsubsection{Supervised training Phase}\label{section: training phase MRA}

\begin{figure}
\centering
\begin{subfigure}{1.0\linewidth}
  \centering
  \includegraphics[width=1.0\linewidth]{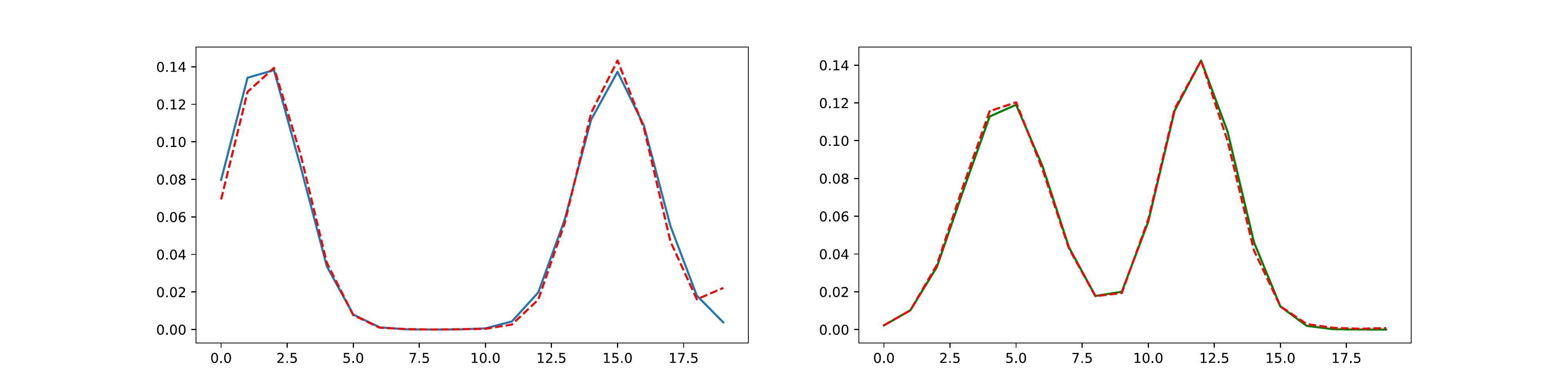}
\end{subfigure}
\begin{subfigure}{1.0\textwidth}
  \centering
  \includegraphics[width=1.0\linewidth]{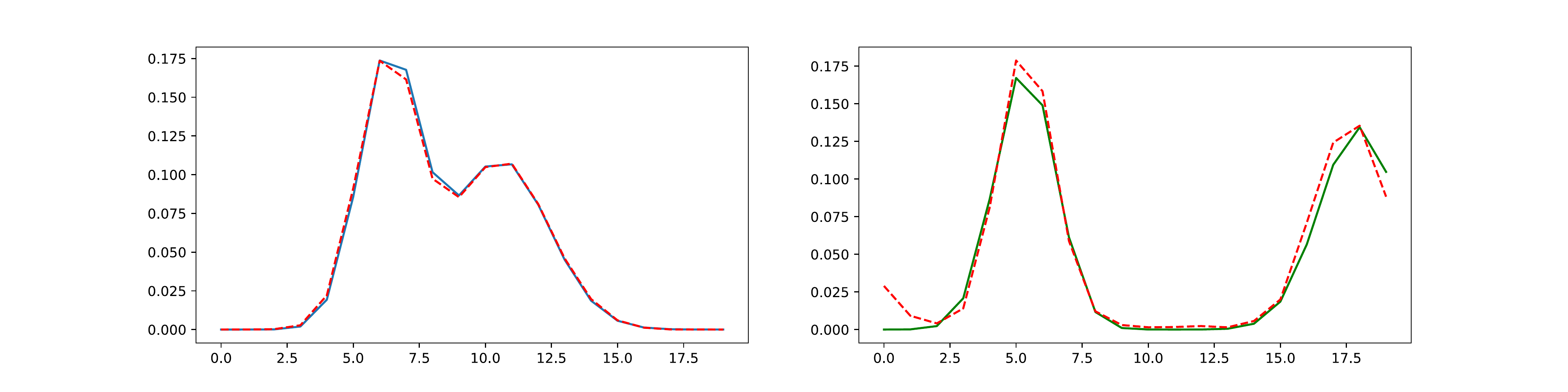}
\end{subfigure}
\begin{subfigure}{1.0\textwidth}
  \centering
  \includegraphics[width=1.0\linewidth]{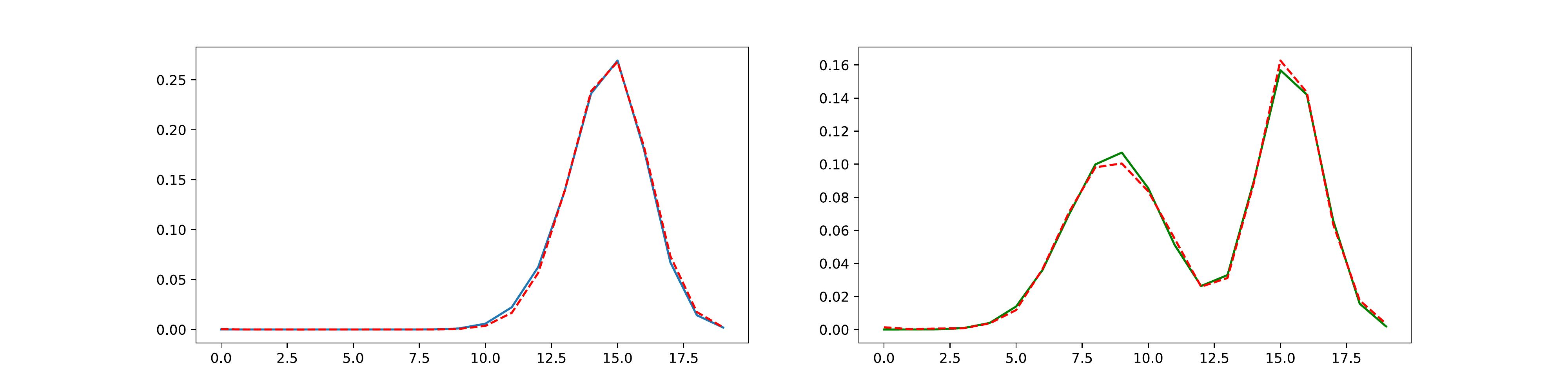}
\end{subfigure}
\caption{Predictions for the distribution $\rho$ (Left) and volume $v$ (Right), outputted by trained encoders $\xi_\theta^\rho$ and $\xi_\theta^v$ respectively, for $\rho,v$ being mixture of $2$ Gaussians. The solid lines are the ground truth $\rho$ and $v$, while the dotted lines are the corresponding predictions by an NN.}
\label{fig:learnt_rho_sig}
\end{figure}

The goal of the training phase is to use the encoder $\xi_\theta$ to predict $\rho(X_1)$ and $v(X_1)$ given moments. In other words, we demonstrate that the moment inversion map can be learned by neural networks in a supervised way. 

To this end, we draw the signal $v$ and the density $\rho$ from a distribution. After forming their corresponding first and second moments, we train our encoder $\xi_\theta$ in a supervised way to take inputs of the form $\left(M_F^1[\widehat v, \rho](K_1), M_F^2[\widehat v, \rho](K_1, K_1)\right)$ and output $\left(\rho (X_1),\widehat v(K_1)\right)$. In our experiments, we let the distribution  of $v$ and $\rho$ be a mixture of Gaussians on the interval $\mathcal{I}$, where we repeat our training procedure separately for a different number of Gaussians. We take $1750k$ of input-output pairs to do the training. We compute test error on $250k$ of samples using \eqref{eq:relative error v mra}.

We now discuss the hyperparameters for training. We train the encoders $\xi_\theta^\rho$ and $\xi_\theta^v$ separately; let us consider $\xi_\theta^\rho$. We take the training set and feed the moments pairs $\left(M_F^1[\widehat v, \rho](K_1), M_F^2[\widehat v, \rho](K_1, K_1)\right)$ to $\xi_\theta^\rho$, which outputs corresponding $z_\rho$ for each pair as a prediction for $\rho (X_1)$. We train $\xi_\theta^\rho$ over a total of $30,000$ epochs with learning rates of $10^{-4}, 10^{-5}$ and $10^{-6}$ over $10,000$ epochs successively.
We then repeat the same process for $\xi_\theta^v$.

Table \ref{table:mixgauss errors} summarizes the average relative error on the training and test sets, using~\eqref{eq:relative error v mra}, while evaluating our trained encoders on mixtures of different numbers of Gaussians. The left and right columns of Figure \ref{fig:learnt_rho_sig} show some comparisons of the encoder output $(z_\rho,z_v)$ with ground truth $(\rho(X_1),\widehat v(K_1))$ during prediction time. 
\begin{table}[ht!]
\begin{center}
\begin{tabular}{c | c c c c} 
 \hline\hline
 No. of Gaussians & $z_\rho$ (Train error) & $z_\rho$ (Test error) & $z_v$ (Train error) & $z_v$ (Test error) \\ [0.5ex] 
 \hline\hline
 1 & 0.042 & 0.048 & 0.048 & 0.052\\ 
 \hline
 2 & 0.121 & 0.141 & 0.156 & 0.170\\
 \hline
 3 & 0.177 & 0.195 & 0.180 & 0.206\\
 \hline
\end{tabular}
\caption{Average reconstruction errors (defined in~\eqref{eq:relative error v mra}) of predictions $z_\rho$ and $z_v$ on training and test sets for mixtures of Gaussians.}
\label{table:mixgauss errors}
\end{center}
\end{table}

\subsubsection{Reconstruction Phase}\label{section: recon phase MRA}
In the previous section, we discussed our process of training the encoder $\xi_\theta$ in a supervised way such that it learns the moment inversion map. A useful application of this trained encoder is when supplied with new, possibly noisy, moments $(\hat M_F^1, \hat M_F^2)$, we can use its outputs as a good initialization for further refinement. In this section, we demonstrate that this procedure leads to faster convergence.

We first talk about how we obtain the estimators $\hat M_F^1, \hat M_F^2$ from observations of the form 
\begin{equation}\label{eqn:real obs}
v_j = s_j\circ v(X_1) + \epsilon_j,\quad j=1,\ldots,N
\end{equation}
where $\epsilon_j\sim N(0,\sigma^2 I_n)$. Let $F\in \mathbb{C}^{n\times n}$ again be the Fourier matrix, we form unbiased moment estimators of the form
\begin{equation}\label{eqn:unbiased estimators}
\hat M_F^1 = \frac{1}{N}\sum_{j=1}^N Fv_j,\quad \hat M_F^2\ =\ \frac{1}{N} \sum_{j=1}^{N}  (F v_j)(Fv_j)^*\ -\ \sigma^2 I_n
\end{equation}
by subtracting a constant term on the diagonal of the empirical second moment. These are used as input to the trained encoder $\xi_\theta$ for prediction.

 Notice that the solution to the MRA problem has a global translation ambiguity. Therefore, it is possible for the encoders $\xi_\theta^v, \xi_\theta^\rho$, to output an approximation to signal $v$ and density $\rho$ up to some arbitrary translations. While this is not a big issue if the predicted signal is all we want, it becomes an issue if we want to refine the predictions further. More precisely, before deploying the encoder for refinement with new incoming moments $\hat M_F^1, \hat M_F^2$, we conduct an alignment to ensure that the outputs $z_\rho = \xi_\theta^\rho(\hat M_F^1, \hat M_F^2)$ and $z_v = \xi_\theta^v(\hat M_F^1, \hat M_F^2)$, upon forming 
\begin{eqnarray}\label{eqn:recon moments}
 M_F^1[z_v, z_\rho](K_1) & = & \sum_{j=1}^n z_\rho(j) \exp(-iK_1s(j))\odot z_v ,\cr
    M_F^2[z_v, z_\rho](K_1, K_1) &=& \sum_{j=1}^n z_\rho(j) \left(\exp(-iK_1s(j))\odot z_v\right)  \left(\exp(-iK_1s(j)) \odot z_v\right)^*,
\end{eqnarray}
matches the inputs $(\hat M_F^1, \hat M_F^2)$ of the encoder. Here by abuse of notation, we treat $z_v,z_\rho$ as a continuous object and apply the functionals $M_F^1, M_F^2$ to them. With an alignment, we can make sure the initial loss
\begin{equation}\label{eqn:mra_loss}
    \mathcal{L}_\textnormal{recon}\ =\ \norm{\hat M_F^1\ - M_F^1[z_v, z_\rho](K_1)}_F\ +\ \lambda \norm{\hat M_F^2\ -\ M_F^2[z_v, z_\rho](K_1, K_1)}_F,
\end{equation}
is small. Recall that $z_v = \xi_\theta^v(\hat M_F^1, \hat M_F^2), z_\rho = \xi_\theta^\rho(\hat M_F^1, \hat M_F^2)$, we further optimize the NN parameters $\theta$ to refine $z_v, z_\rho$ with the loss in \eqref{eqn:mra_loss}.

\begin{figure}[ht!]
\centering
\begin{subfigure}{.48\textwidth}
  \centering
  \includegraphics[width=1.0\linewidth]{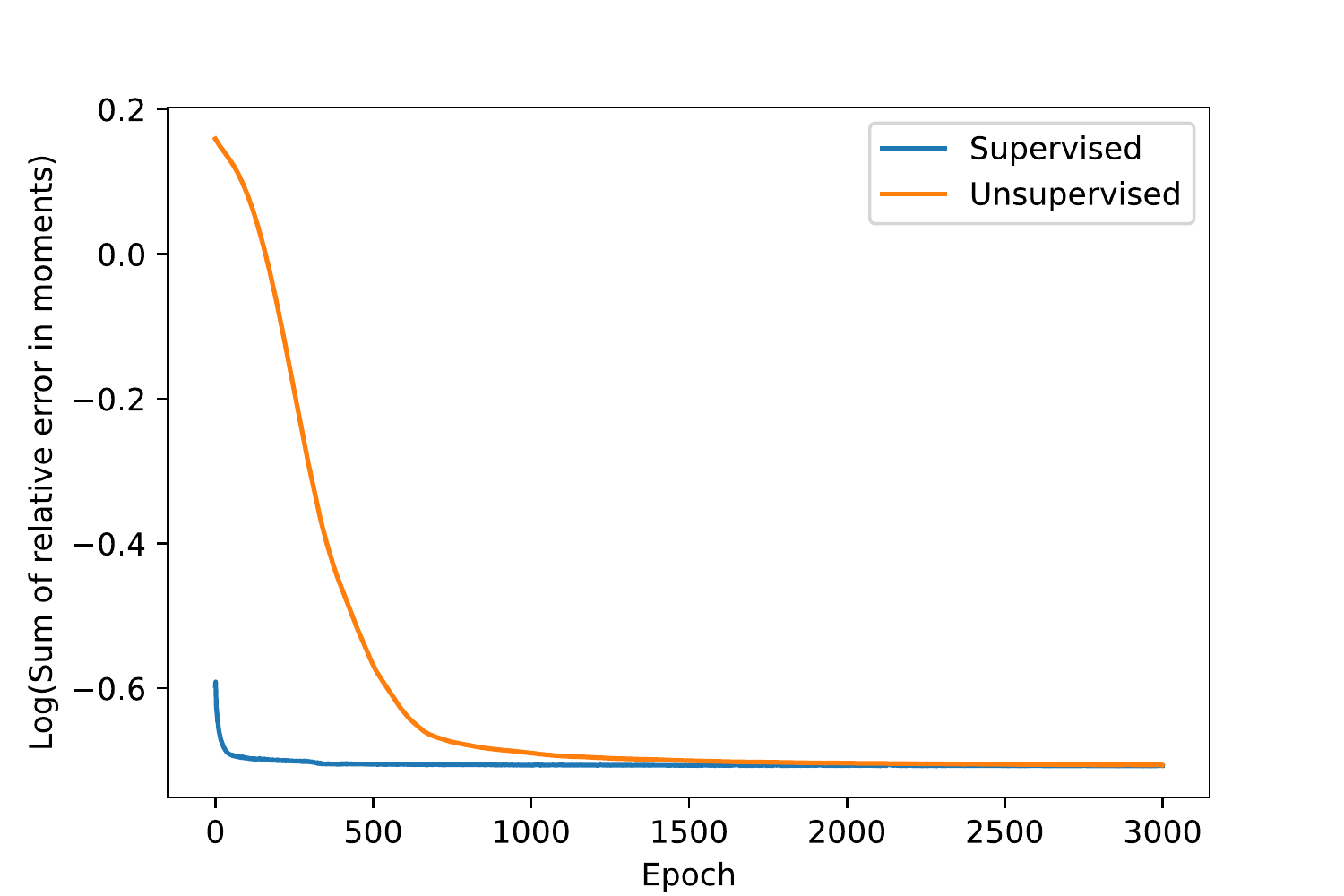}
\end{subfigure}
\begin{subfigure}{.48\textwidth}
  \centering
  \includegraphics[width=1.0\linewidth]{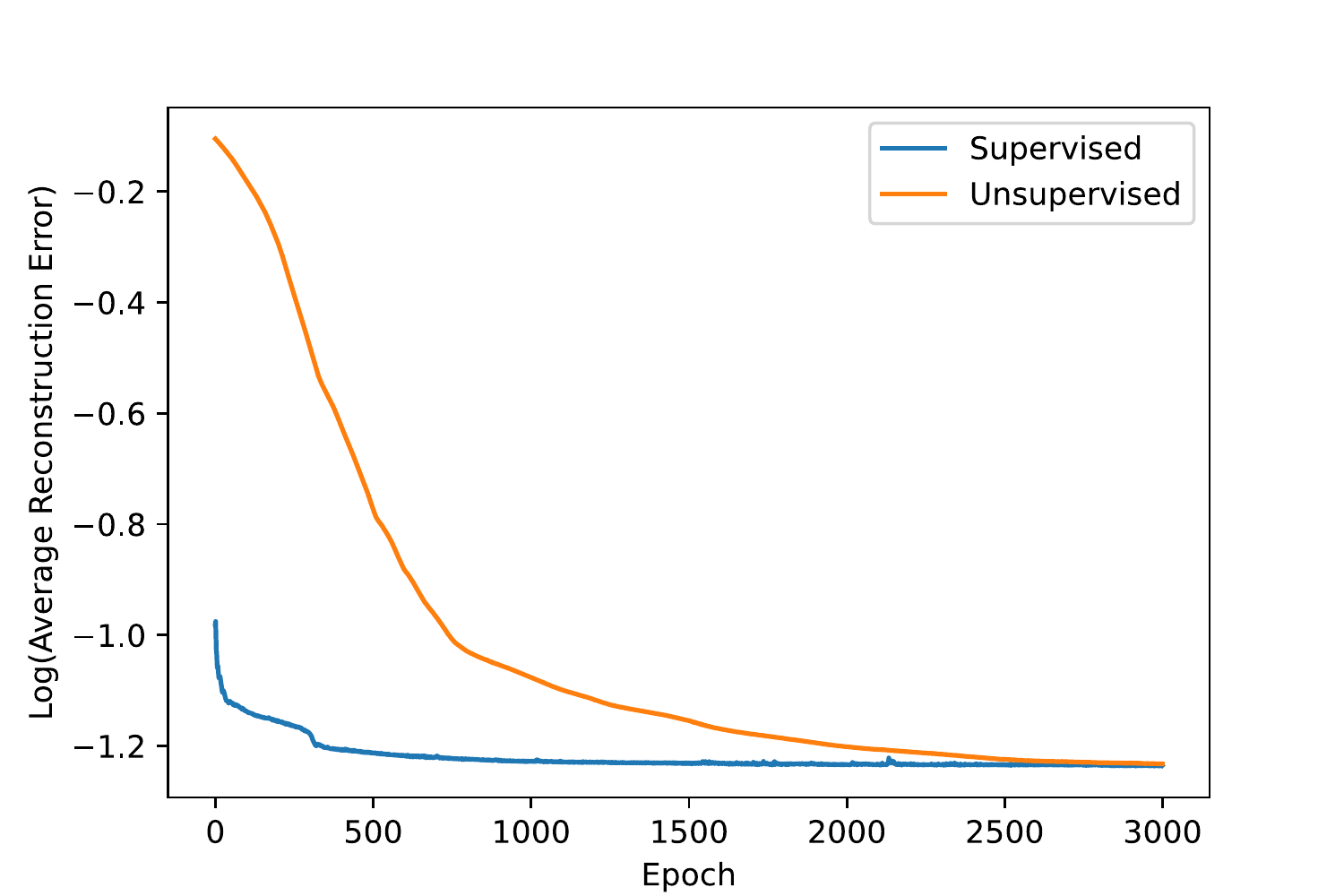}
\end{subfigure}
\caption{Plots of logarithms (with base $10$) of Sum of relative errors (defined in~\eqref{eq:relative error moments}) for $\hat M_F^1$ and $\hat M_F^2$ across 3000 iterations (Top), and Reconstruction error (defined in~\eqref{eq:relative error v mra}) across 3000 iterations (Bottom); averaged over $20$ reconstructions of $(\rho(X_1),\widehat v(K_1))$ pairs drawn from the family of a mixture of $2$ Gaussians. In both plots, the blue curve corresponds to the scenario where the encoder underwent supervised training, while the orange corresponds to the scenario where it did not.}
\label{fig:learn_rho_sig_plot}
\end{figure}

We now show the results of the deployment of our architecture $\xi_\theta$ when working with noisy moments. We take 20 different $\hat M_F^1, \hat M_F^2$, and determine $z_v,z_\rho$ by minimizing \eqref{eqn:mra_loss} over the parameters of $\xi_\theta$. The relative errors (defined in~\eqref{eq:relative error v mra} and \eqref{eq:relative error moments}) of the reconstructed $(\rho(X_1),\widehat v(K_1))$ and the moments are plotted in Figure \ref{fig:learn_rho_sig_plot}. The errors are averaged over $20$ different instances of $(\rho, v)$ combinations from mixtures of $2$ Gaussians,  and the empirical moments are formed from 1000k observations for each pair of $(\rho, v)$ as in~\eqref{eqn:real obs}, with Gaussian noise $\sigma = 1.0$.  Depending on whether the encoder underwent supervised training, we observe the trajectory of this ``average'' reconstruction error to be different. Figure \ref{fig:learn_rho_sig_plot} illustrates that the average reconstruction error indeed converges faster when the encoder is trained in a supervised phase. 

\subsection{Cryo-EM}

\begin{figure}
\centering
\begin{subfigure}{.48\linewidth}
\centering
\includegraphics[width=1.0\linewidth]{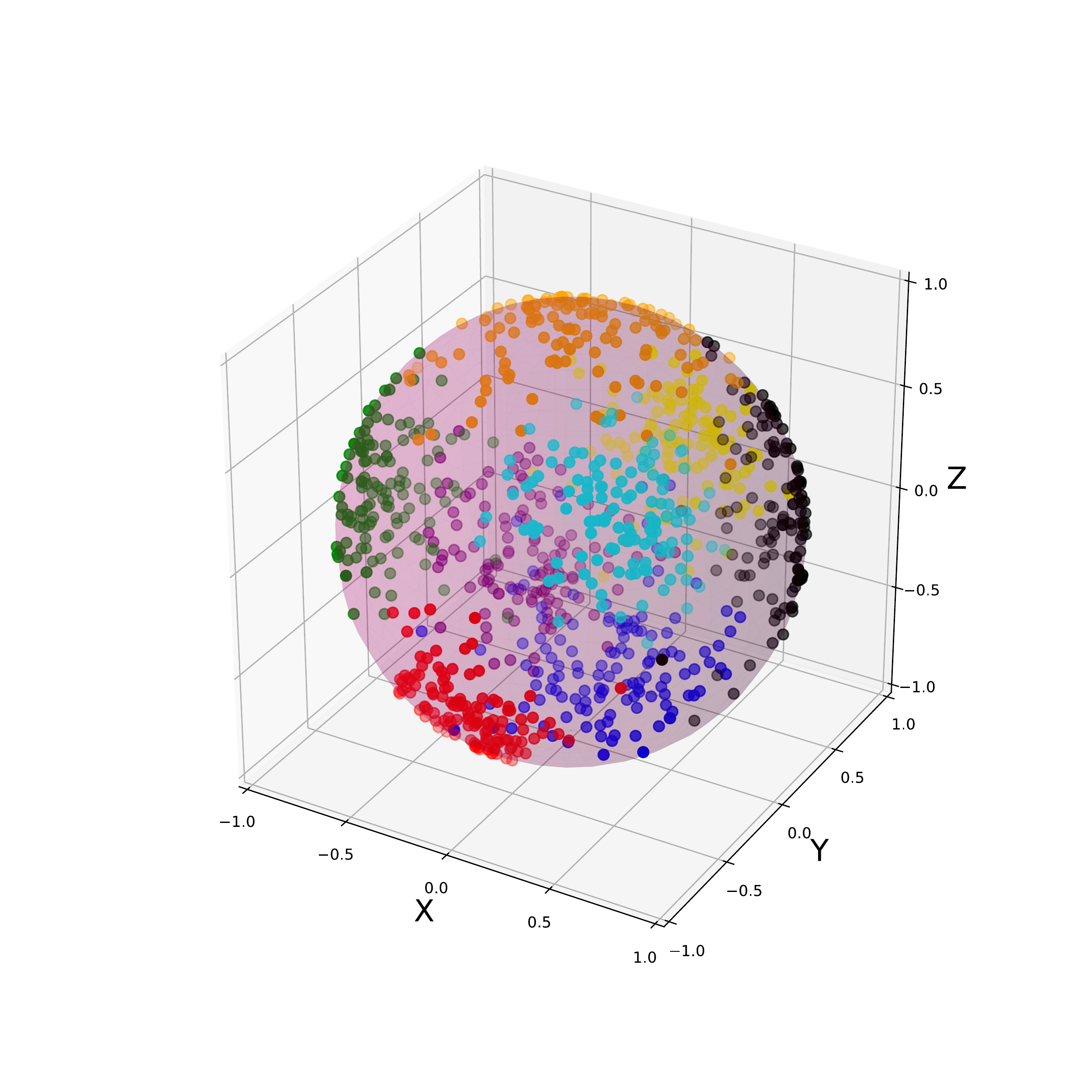}
\end{subfigure}
\begin{subfigure}{.48\textwidth}
\centering
\includegraphics[width=1.0\linewidth]{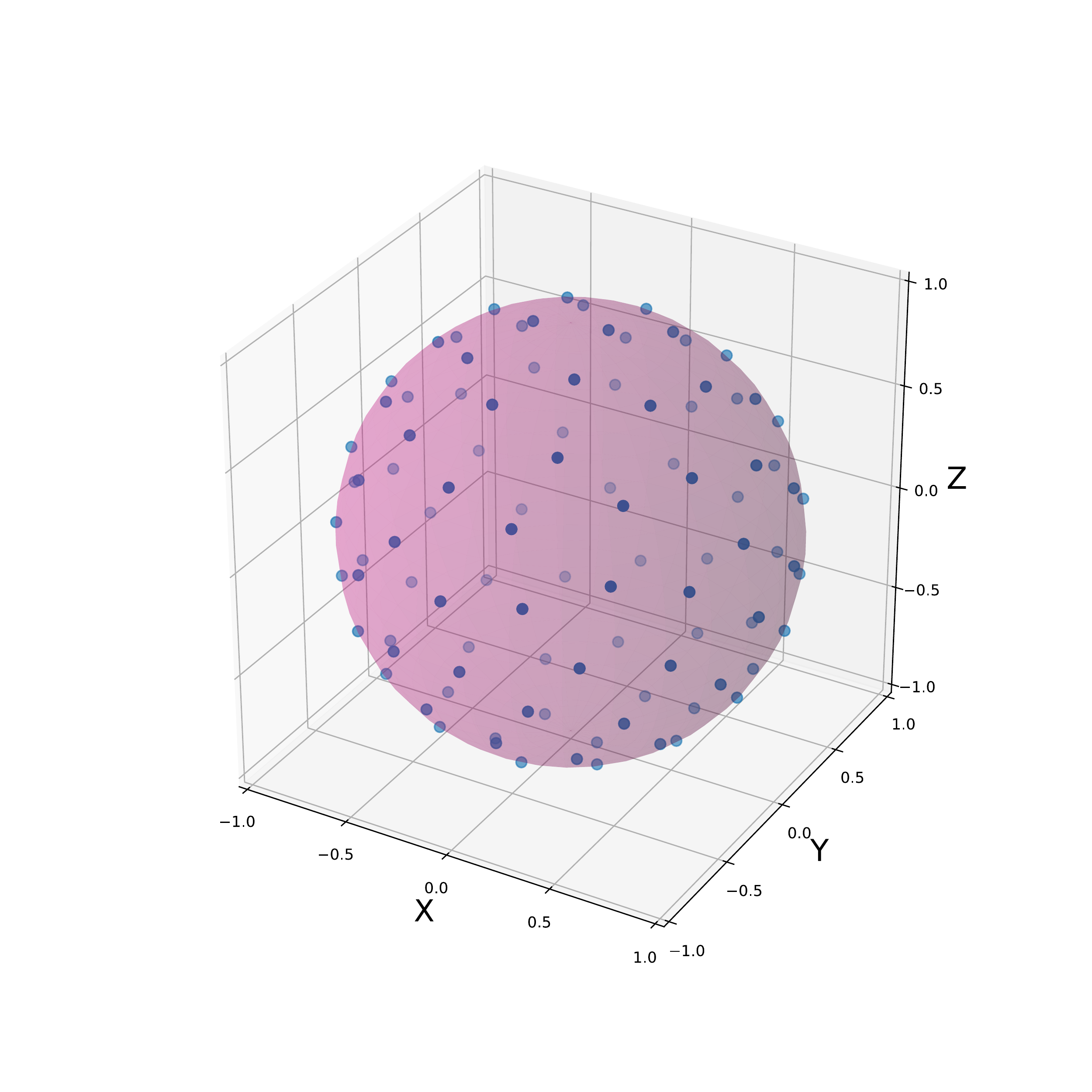}
\end{subfigure}
\caption{(Left) 1000 points sampled from a mixture of eight von Mises-Fisher random variables shown in different colors, and (Right) $100$-point $13$-design plotted on a 3D unit sphere.}
\label{fig:vonMisesFisher}
\end{figure}

We now present the results using our method for cryo-EM as illustrated in Section \ref{sec:amortized_cryo}. Again for evaluation purposes, the relative error for an estimate $u\in\mathbb{R}^{n^3}$ of a signal $v$ discretized at $n^3$ equispaced points $X_3$ on $\mathcal{I}^3$, is defined as
\begin{equation}\label{eq:relative error v cryo}
    \inf_{R\in \so{3}} \frac{\norm{R\circ v\ \left(X_3\right)\ -\ u}_F}{\norm{v \left(X_3\right)}_F}.
\end{equation}
The relative errors for moment estimators of the first and second moments are defined analogously to~\eqref{eq:relative error moments}.

While we do not describe any supervised training phase like in the MRA case, our architecture keeps this option open. We believe that even for cryo-EM, it would be possible to train our encoder $\xi_\theta^\rho$ in a supervised way to learn the moment inversion map, i.e., to take inputs of the form $\left(M_F^1[\widehat v,\rho](K_2), M_F^2[\widehat v,\rho](K_2,K_2)\right)$ and predict $\left(\rho(R) \right)_{R \in Q}$ for training and reconstruction, where $Q$ is the set of quadrature points on $\so{3}$ defined in \ref{sec:amortized_cryo}. It would also be possible to train $\xi_\theta^v$ such that it outputs a discretized approximation of the volume from the moments, or at least some vector containing important feature information about it.

The reconstruction is carried out by optimizing the NN parameters $\theta$ and $\phi$ of our encoder $z_\rho=\xi_\theta^\rho(\hat M_F^1,\hat M_F^2)$ and neural representation $\widehat v_\phi$, respectively, to minimize the loss function
\begin{equation}\label{eqn:cryo_loss}
    \mathcal{L}_\textnormal{recon}\ =\ \norm{\hat M_F^1\ - M_F^1[\widehat v_\phi, z_\rho](K_2)}_F\ +\ \lambda \norm{\hat M_F^2\ -\ M_F^2[\widehat v_\phi, z_\rho](K_2, K_2)}_F.
\end{equation}
During reconstruction, one of the challenges we face is fixing a good set $Q\subset \so{3}$ on which we shall use a quadrature rule with uniform weights to evaluate the functionals $M_F^1, M_F^2$, as described in~\eqref{eqn:quadrature_moments}. In our experiments, we do so in two steps. First, we choose a $q_1$-point spherical design on $S^2$, see, e.g.,~\cite{womersley2018efficient}. A $q_1$-point spherical $t$-design is a finite set of points with cardinality $q_1$ on $S^2$, such that their quadrature over $S^2$ with uniform unit weights is exact for any polynomial (spherical harmonics) with degree $\leq t$. Then, for each point of the design, treating the axis connecting that point to the center as viewing direction, we consider in-plane rotations with $q_2$ equally spaced angles in $[0, 2\pi)$ radians. This gives us a set $Q$ with $|Q| = q_1 q_2$ quadrature points on $\so{3}$. In experiments, we take $q_1 = 100$ and $q_2 = 12$ for a total of $|Q| = 1200$ quadrature points. To illustrate these quadrature points, we use a $100$-point $13$-design on $S^2$ as the set of viewing directions, as seen in the right side of Figure \ref{fig:vonMisesFisher}.

\begin{figure}
\centering
\includegraphics[width=0.45\textwidth]{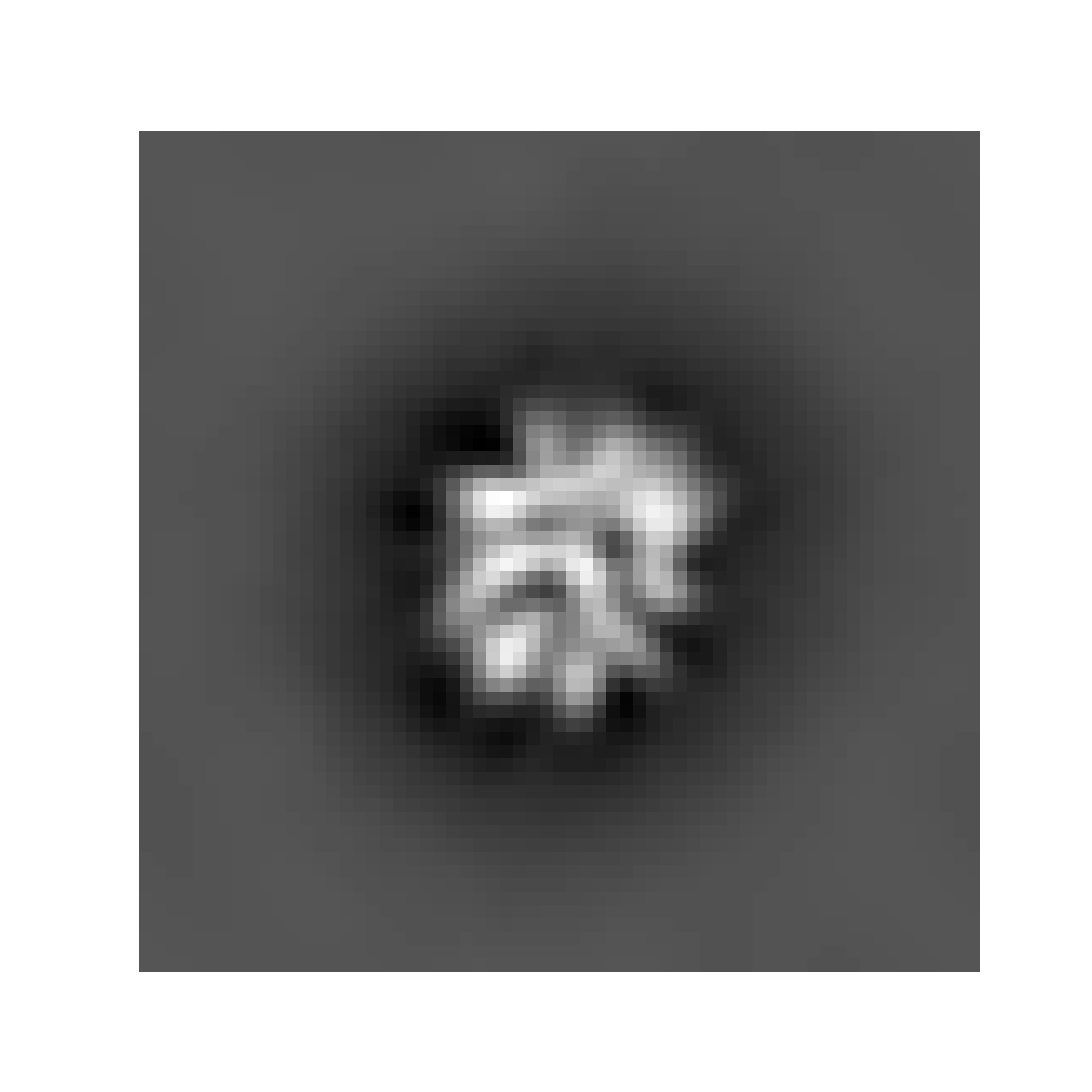}
\includegraphics[width=0.45\textwidth]{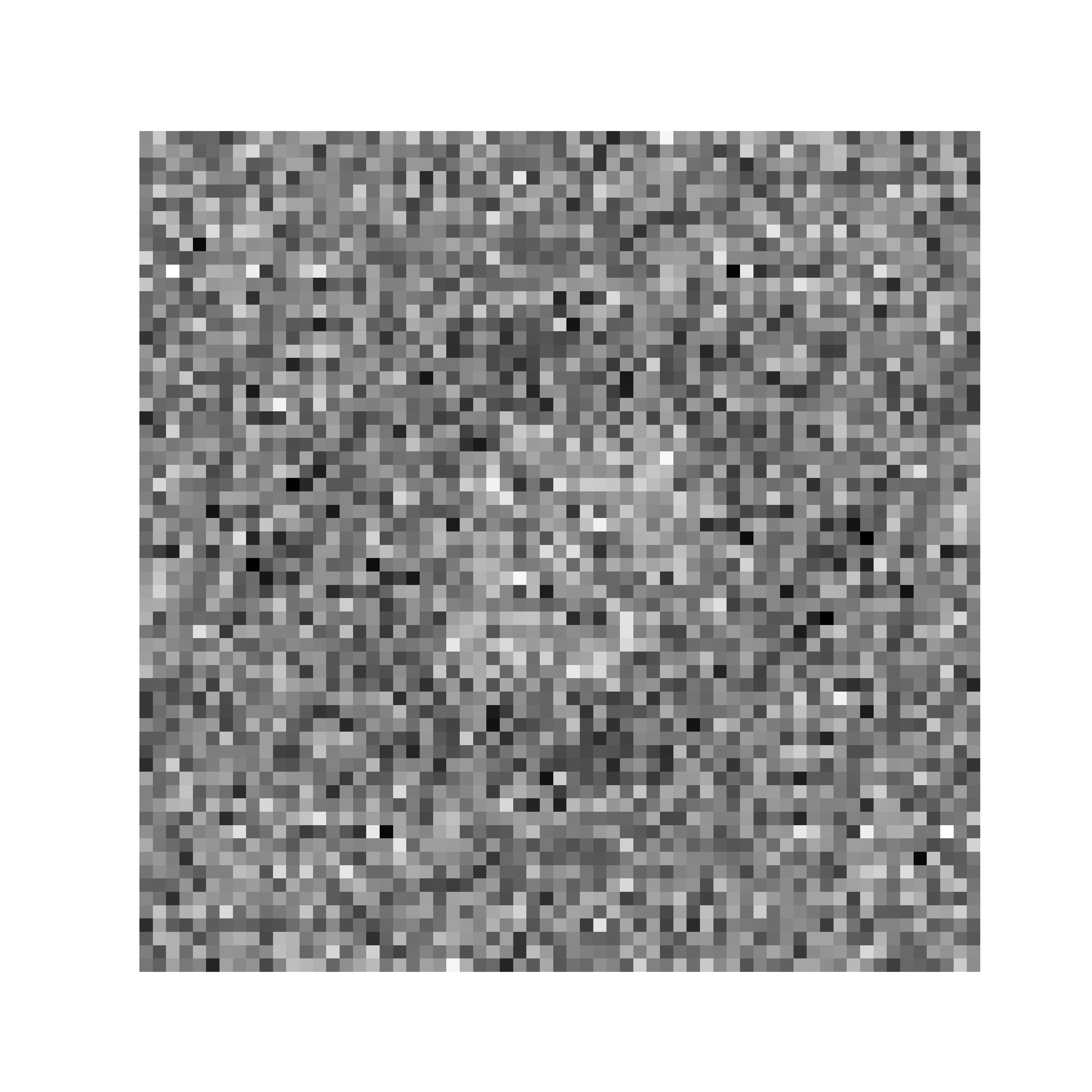}
\caption{(Left) A clean projection, and (Right) its noisy counterpart with noise level $\sigma = 0.5$ as defined in~\eqref{eqn:real obs2}, for EMD-$0409$}
\label{fig:clean vs noisy othermol}
\end{figure}

\begin{figure}
\centering
\includegraphics[width=0.45\textwidth]{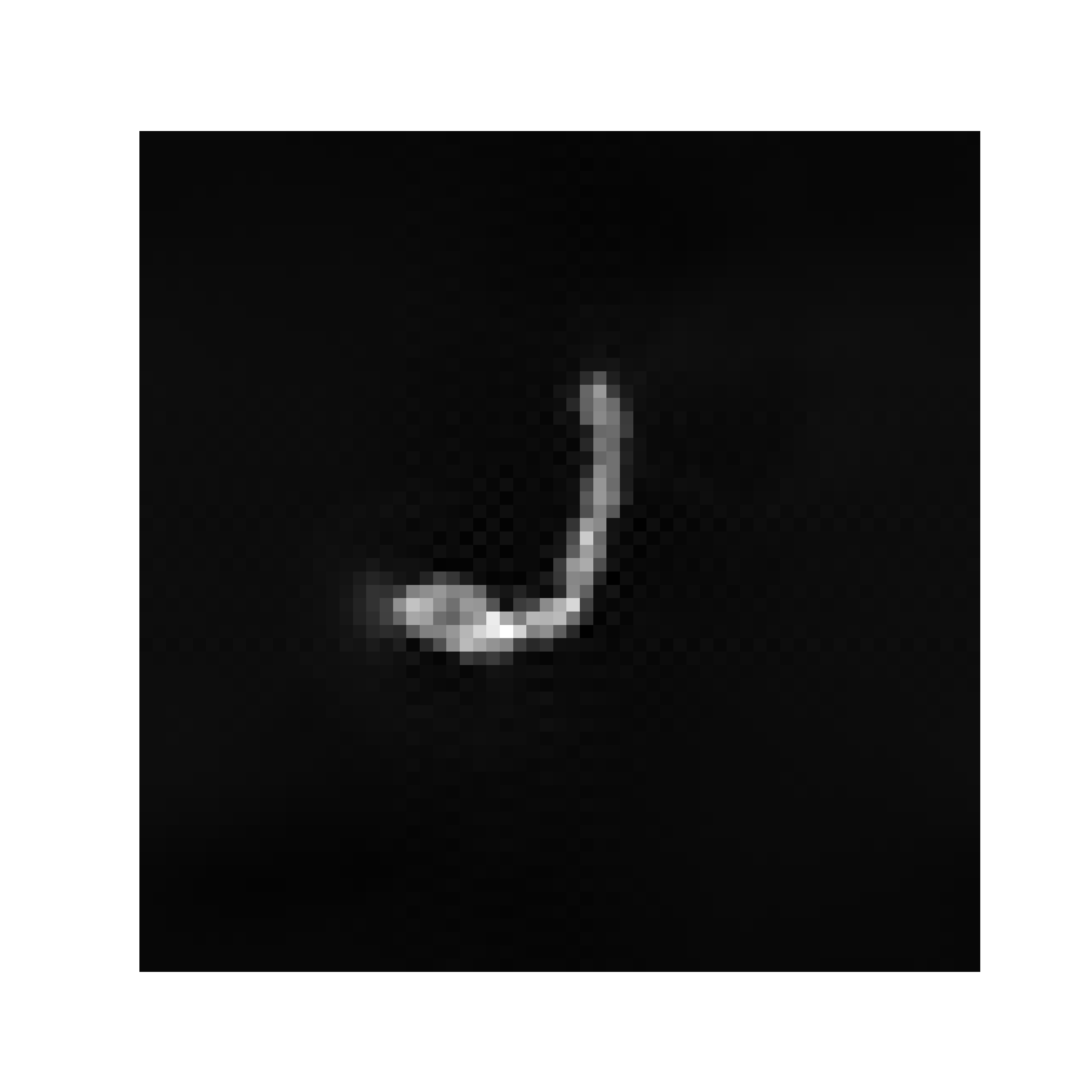}
\includegraphics[width=0.45\textwidth]{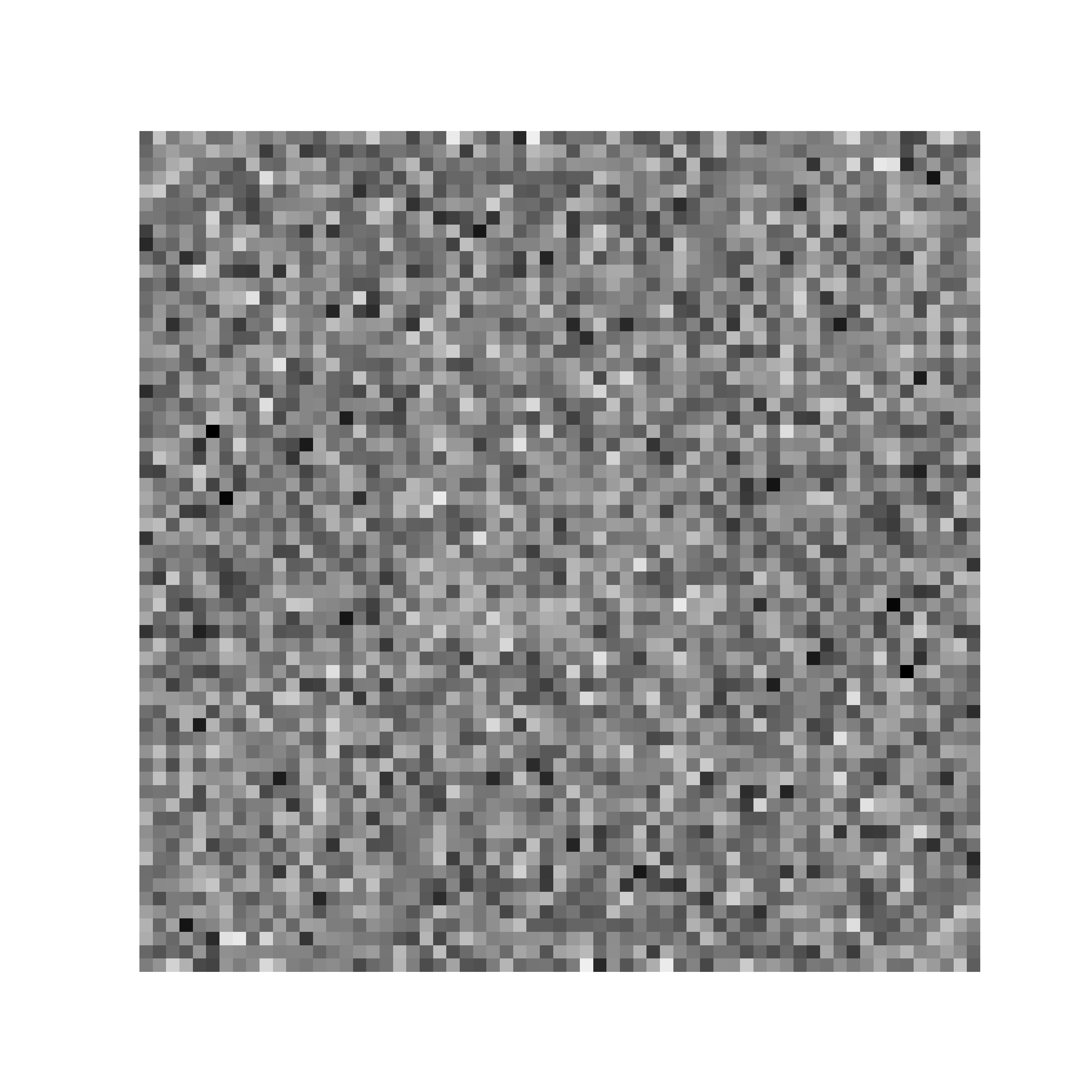}
\caption{(Left) A clean projection, and (Right) its noisy counterpart with noise level $\sigma = 0.5$ as defined in~\eqref{eqn:real obs2}, for EMD-$25892$}
\label{fig:clean vs noisy banana}
\end{figure}

We now discuss our data generation process for cryo-EM and the moment estimators to be inputted into the encoders. In practice, given real observations of the form 
\begin{equation}\label{eqn:real obs2}
v_j = \mathcal{P}\circ R_j\circ v (X_2) + \epsilon_j,\quad j=1,\ldots,N
\end{equation}
where $\epsilon_j\sim N(0,\sigma^2 I_{n^2})$ and $X_2$ is $n^2$ equispaced points on $\mathcal{I}^2$, we could form unbiased moment estimators
\begin{equation}\label{eqn:real obs estimators}
\hat M_F^1 = \frac{1}{N}\sum_{j=1}^N F_2 v_j,\quad \hat M_F^2\ =\ \frac{1}{N} \sum_{j=1}^{N}  (F_2 v_j)\otimes (F_2 v_j)\ -\ \sigma^2 I_{n^2},
\end{equation}
letting $F_2\in \mathbb{C}^{n^2\times n^2}$ be the two-dimension Fourier transform matrix. Clean observations $v_j$ are depicted alongside their noisy counterparts in Figures \ref{fig:clean vs noisy othermol} and \ref{fig:clean vs noisy banana}.

We next discuss our choices of ground truth volumes $v$ and rotational distributions $\rho$. For our experiments, we use three volumes: EMD-$0409$ and EMD-$25892$ taken from the Electron Microscopy Data Bank (EMDB); and a mixture of four Gaussians not lying on the same plane in three dimensions. The dimensions of EMD-$0409$ are $128\times 128\times 128$ with voxel size $1.117\ \angstrom$, while the dimensions of EMD-$25892$ are $320\times 320\times 320$ with voxel size $1.68\ \angstrom$. Both volumes were downsampled to $63\times 63\times 63$ and scaled to have norm $1$. The mixture of Gaussians has dimensions $25\times 25\times 25$, whose voxel size is taken to be $1\ \angstrom$ since it is a simulated volume. We represent the ground truth using $\widehat v_\phi$, and report the approximation error as defined in~\eqref{eq:relative error v cryo}, between the original and this \textit{neural} ground truth as $0.043$ for EMD-$0409$, $0.076$ for EMD-$25892$, and $0.004$ for the mixture of Gaussians. These NN-approximated volumes are then used as the ground truths for the rest of the simulations. The ground truth distribution of rotations $\rho$ is chosen in the following way. The viewing directions are distributed as a mixture of $8$ von Mises-Fisher distributions with different mean directions $\mu$ and concentration parameters $\kappa$, respectively, to ensure a sufficiently non-uniform distribution on $S^2$. $1000$ points from this distribution are shown on the left side of Figure \ref{fig:vonMisesFisher}. The in-plane rotations are uniform on $[0, 2\pi)$ and independent of the viewing directions. We then create moment estimators from $N = 5,000,000$ noisy observations with noise level $\sigma = 0.5$ using~\eqref{eqn:real obs estimators}, where a neural slice approximates $F_2 v_j$.

\begin{figure}
\centering
\includegraphics[width=0.48\textwidth]{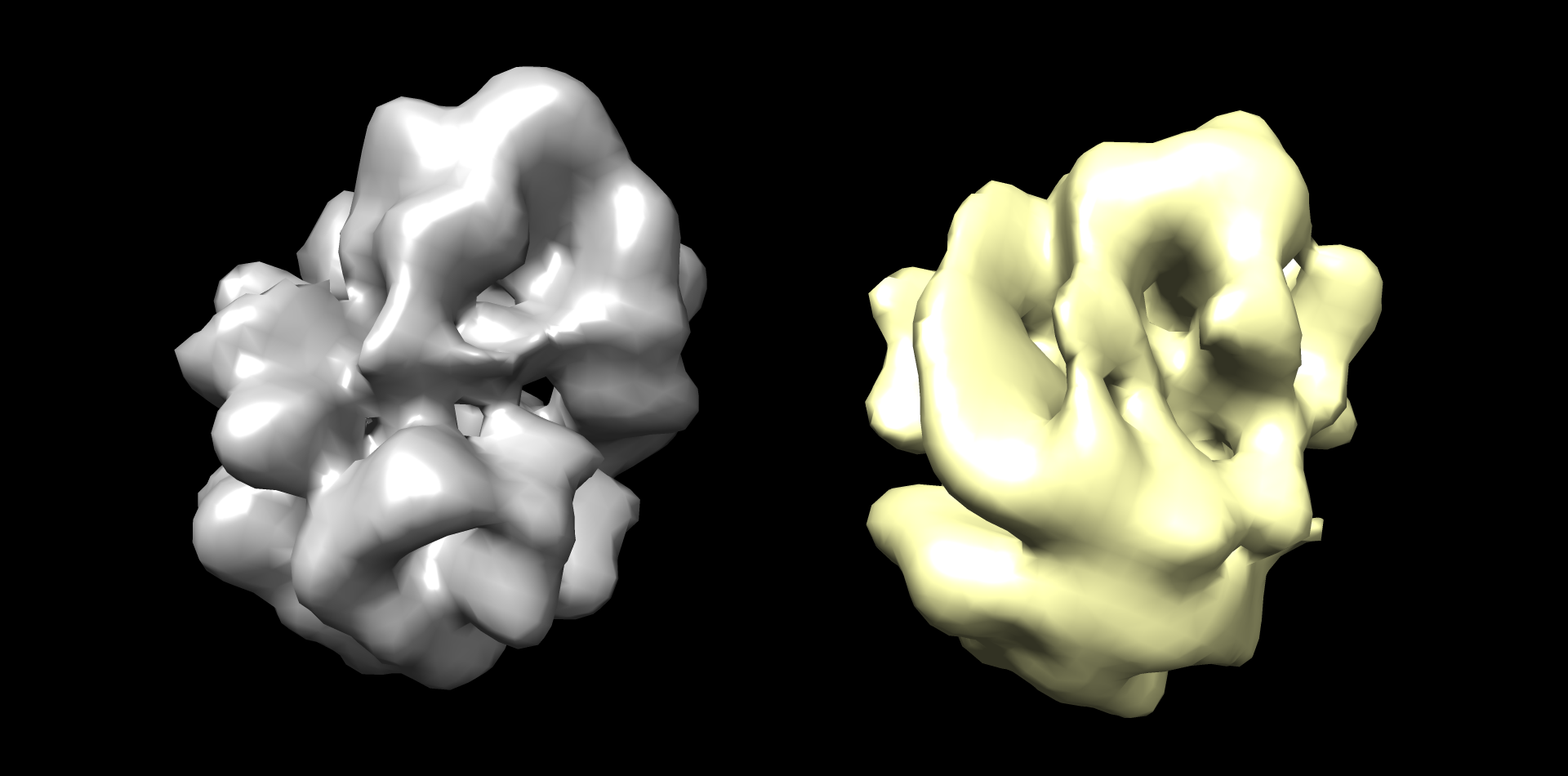}
\includegraphics[width=0.46\textwidth]{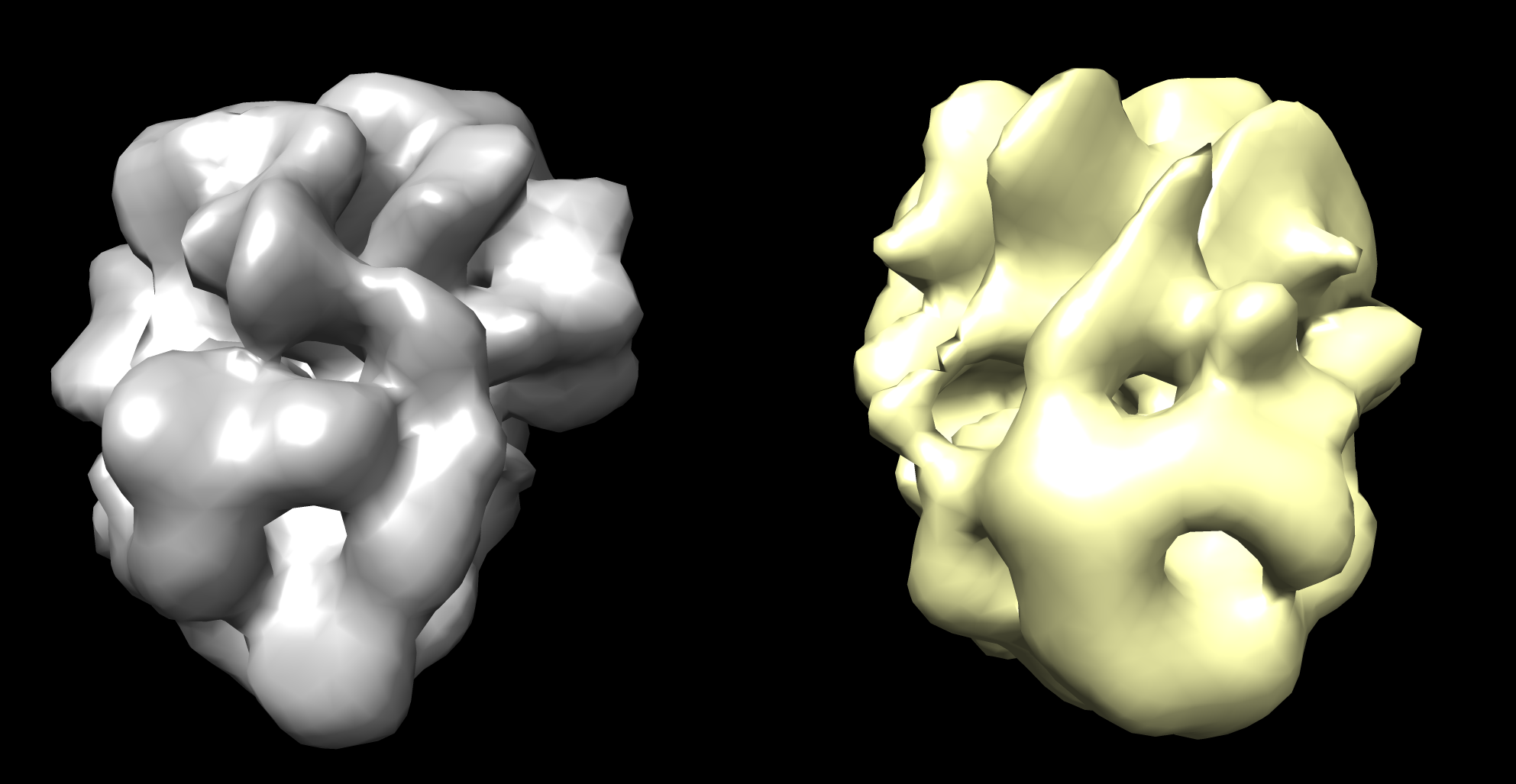}
\caption{Ground truth volume (in gray) and reconstructed volume (in yellow) for the EMD-$0409$ volume, visualized using UCSF Chimera~\cite{pettersen2004ucsf}.}
\label{fig:cryo_results_othermol}
\end{figure}

\begin{figure}
\centering
\includegraphics[width=0.48\textwidth]{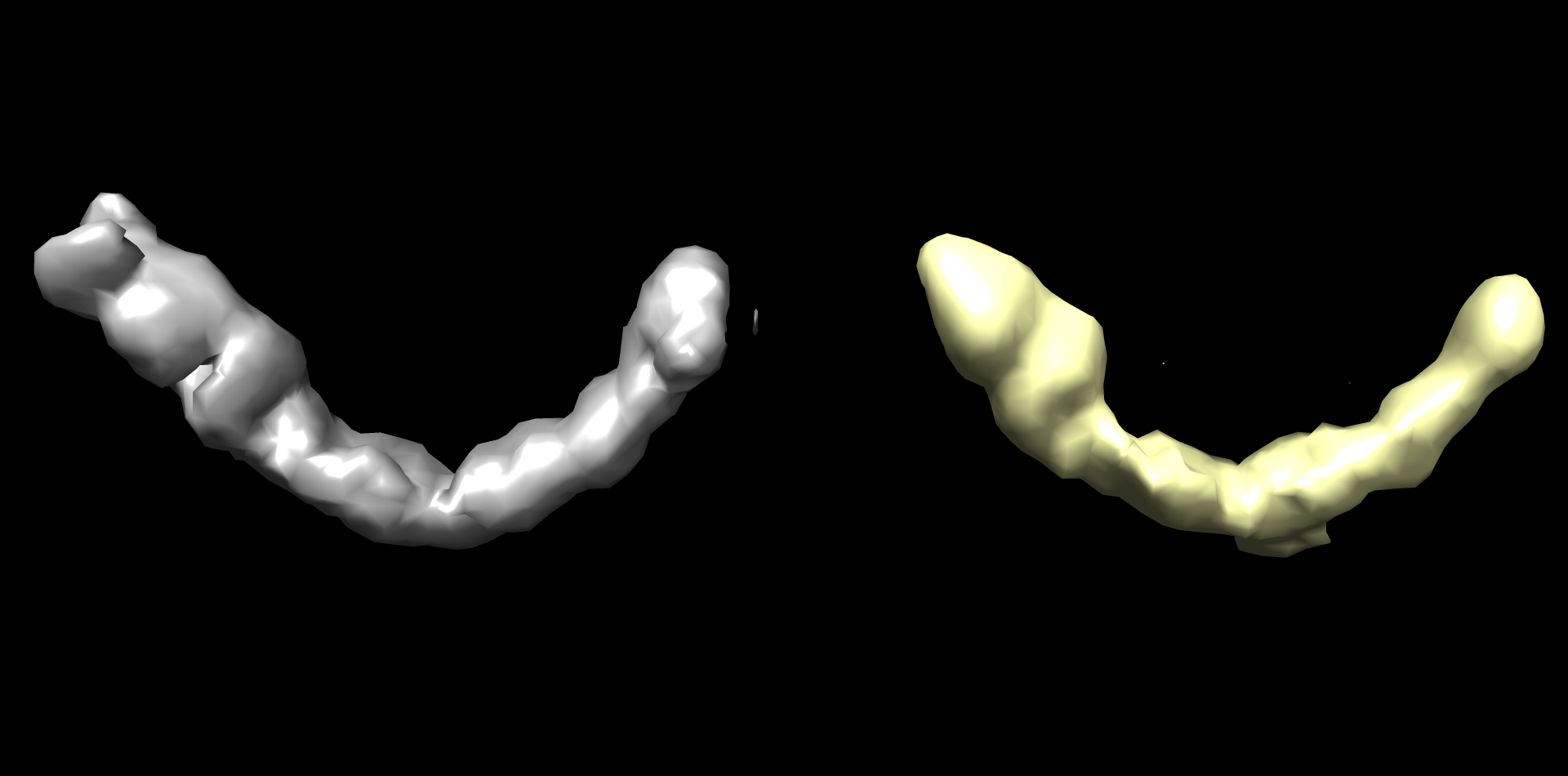}
\includegraphics[width=0.48\textwidth]{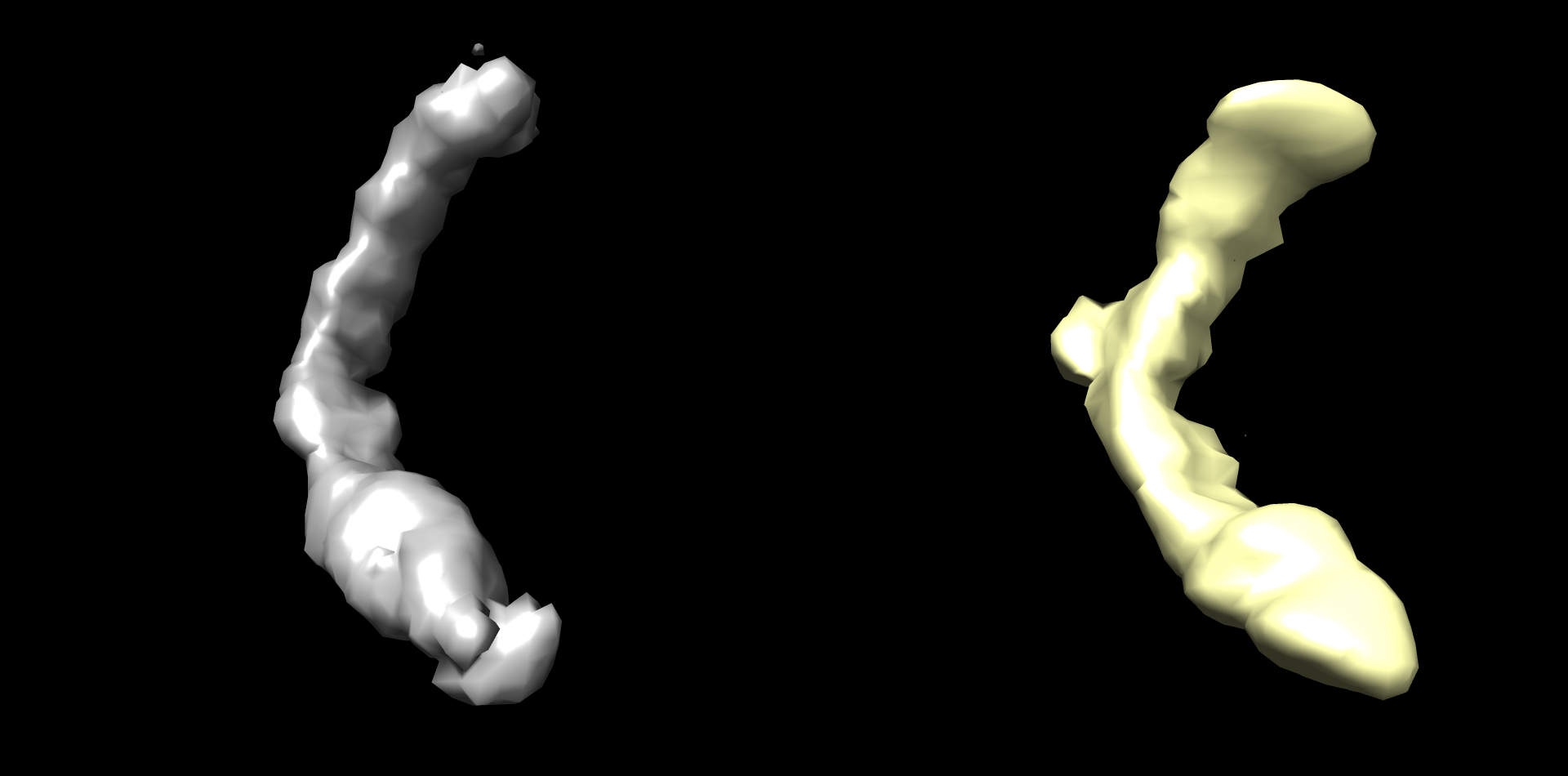}
\caption{Ground truth volume (in gray) and reconstructed volume (in yellow) for the EMD-$25892$ volume, visualized using UCSF Chimera~\cite{pettersen2004ucsf}.}
\label{fig:cryo_results_banana}
\end{figure}

\begin{figure}
\centering
\includegraphics[width=0.48\textwidth]{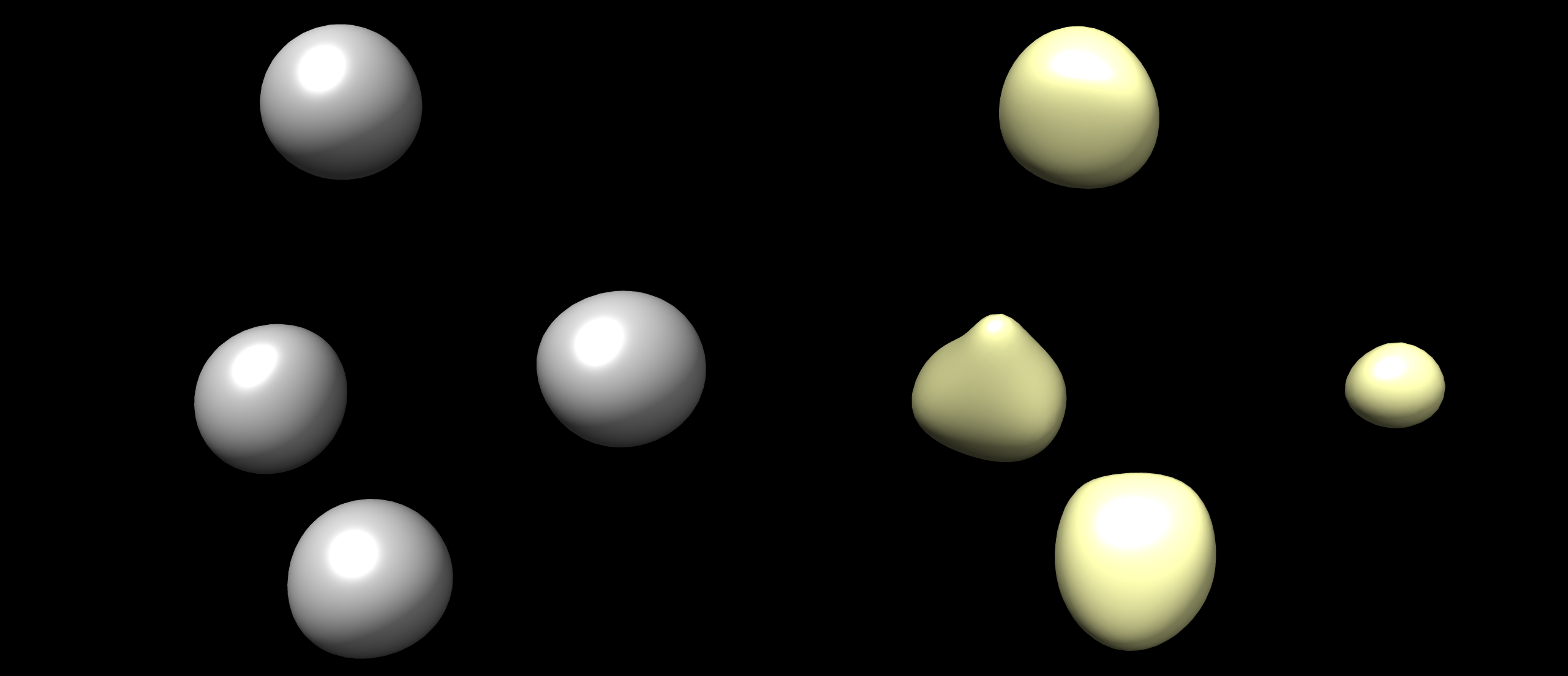}
\includegraphics[width=0.48\textwidth]{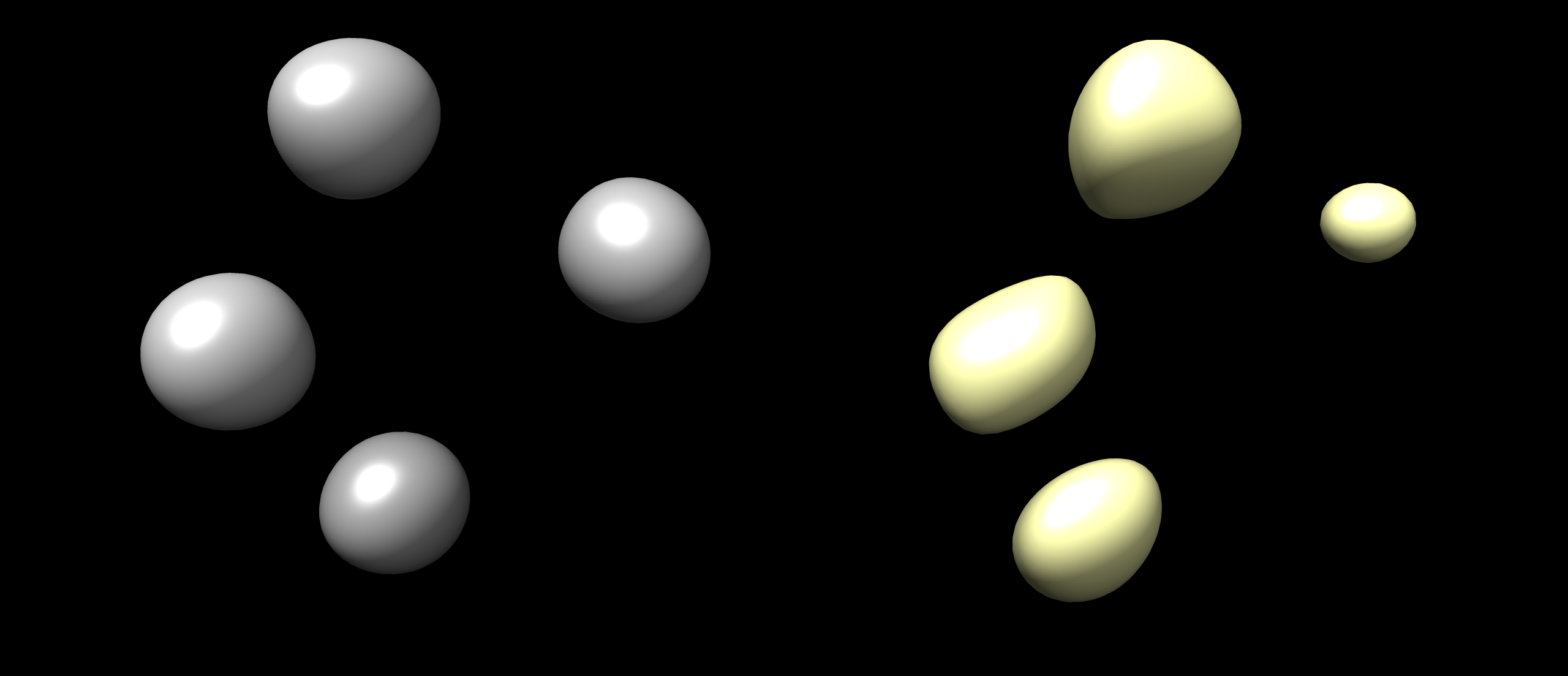} 
\caption{Two views of recovery of a mixture of Gaussians. Ground truth volume (in gray) and reconstructed volume (in yellow) for a mixture of $4$ Gaussians in three dimensions, visualized using UCSF Chimera~\cite{pettersen2004ucsf}.}
\label{fig:cryo_results_synth}
\end{figure}

We run our algorithm with learning rates $10^{-5}$ and $10^{-6}$ successively for $10,000$ epochs each, to minimize the loss function in~\eqref{eqn:cryo_loss}. The reconstructed volumes are visualized in Figures \ref{fig:cryo_results_othermol}, \ref{fig:cryo_results_banana}, and \ref{fig:cryo_results_synth}, alongside their corresponding neural ground truth volumes for EMD-$0409$, EMD-$25892$, and mixture of Gaussian volumes, respectively. Table \ref{table:cryoem errors moments} shows the relative errors of our moments from the reconstructed volumes, defined analogously to~\eqref{eq:relative error moments}, at the end of our reconstruction.

Finally to evaluate the quality of reconstruction, we first align the reconstructed volumes with the ground truth. For that purpose, we run the algorithm for aligning three-dimensional density maps in~\cite{harpaz2023three} multiple times and pick the best alignment. We then calculate the Fourier Shell Correlation (FSC) between the ground truth volumes and their corresponding aligned reconstructions. We denote the resolution of the reconstructed volume as the point where the FSC curve goes below $0.5$. The final resolutions between the ground truths and reconstructed volumes are provided in Table \ref{table:cryoem resolutions}.

\begin{table}[ht!]
\begin{center}
\begin{tabular}{c | c c} 
 \hline\hline
 Volume & Relative error in $\hat M_F^1$ & Relative error in $\hat M_F^2$ \\ [0.5ex] 
 \hline\hline
 EMD-$0409$ & 0.003 & 0.013 \\
 \hline
 EMD-$25892$ & 0.007 & 0.035 \\
 \hline
 Mixture of Gaussians & 0.007 & 0.016 \\ 
 \hline
\end{tabular}
\caption{Final relative errors of moment estimates $\hat M_F^1$ and $\hat M_F^2$ after reconstruction phase.}
\label{table:cryoem errors moments}
\end{center}
\end{table}

\begin{table}[ht!]
\begin{center}
\begin{tabular}{c | c c} 
 \hline\hline
 Volume & Resolution (in $\angstrom$) \\ [0.5ex] 
 \hline\hline
 EMD-$0409$ & 16.86 \\
 \hline
 EMD-$25892$ & 21.52 \\
 \hline
 Mixture of Gaussians & 4.45 \\ 
 \hline
\end{tabular}
\caption{Optimal resolutions between ground truth volumes and their reconstructions.}
\label{table:cryoem resolutions}
\end{center}
\end{table}

\section{Conclusion and outlook}\label{final_thoughts}

\rev{Single-particle cryo-EM is a prominent method for determining the atomic-resolution 3D structure of biological macromolecules. This technique underwent a ``resolution revolution'' a decade ago~\cite{kuhlbrandt2014resolution}, and three of its pioneers were awarded the 2017 Nobel Prize in Chemistry~\cite{dubochet2017nobel}. These days, cryo-EM provides researchers access to some of the molecules' tiniest and most essential building blocks. In this paper, we addressed the reconstruction problem in cryo-EM as well as one of its simpler versions, namely, multirefence alignment. Both cryo-EM and MRA fall under the class of orbit recovery problems.}

Although deep NN-based methods have been successfully used in maximum likelihood estimation for orbit recovery problems, they have not historically exploited the benefits offered by the MoM, like noise resilience, due to the central limit theorem when averaging data. In this paper, we take a first step towards using NNs for solving moment systems in orbit recovery problems. In the case of MRA, we demonstrate theoretically and numerically that a map can be learned to take moments as input and output the signal and density of translations, \rev{and develop novel neural network architectures for the same}. This map can then be used as a deep NN prior to accelerating convergence in unsupervised reconstruction from new incoming moments.

We also apply this approach to cryo-EM with encouraging results, but further work is needed to demonstrate the superiority of supervised learning and tackle more general cryo-EM models, \rev{like those dealing with small translations in addition to the rotations, and further image contamination due to aberrations (which would involve accounting for contrast transfer functions). Supervised learning would effectively enable low-dimension reconstruction of volumes near-instantly and would serve as an inexpensive and time-efficient method of generating \textit{ab-initio} models for iterative refinement algorithms.} Other future work includes investigating the use of higher-order moments to improve reconstruction accuracy and parallelizing the model on multiple GPUs to enable reconstruction with larger images and improve speed and accuracy. Additionally, tackling more general cryo-EM models will bring us closer to operating on real-world datasets.

\section*{Acknowledgements}
YK is thankful to DOE for funding DE-SC$0022232$. NS is partially supported by the NSF-BSF award $2019752$ and DFG award $514588180$.

\bibliographystyle{plain}
\bibliography{A_MoM_bib}

\appendix

\section{Architecture of neural networks}\label{architecture}

In this appendix, we describe the details of the NN for both MRA and cryo-EM. To facilitate the discussion, we first define $\texttt{conv1D}_{w,c}$ to be a 1D convolutional layer with periodic padding, kernel window size $w$ and channel number $c$. In a similar way, we also denote a 2D convolutional layer with window size $w\times w$ and channel number $c$ as $\texttt{conv2D}_{w,c}$. Furthermore we define $\texttt{input1D}_{\ell,c}$ to be an input layer that prepares the input as a length $\ell$ 1D vector field with channel number $c$. We then define a fully connected layer $\texttt{full}_{w}$ that takes an input vector field and output a vector with size $w$. The nonlinearities we use in this paper are leaky ReLu (LReLu) nonlinear activation with parameter 0.02, $\tanh(\cdot)$ function, and just linear activation (without nonlinearities). We make no distinction between real or complex input, since changing real to complex input only requires doubling the input or output channel number. 

\subsection{MRA}
 
\begin{figure}
    \includegraphics[width = 15cm]{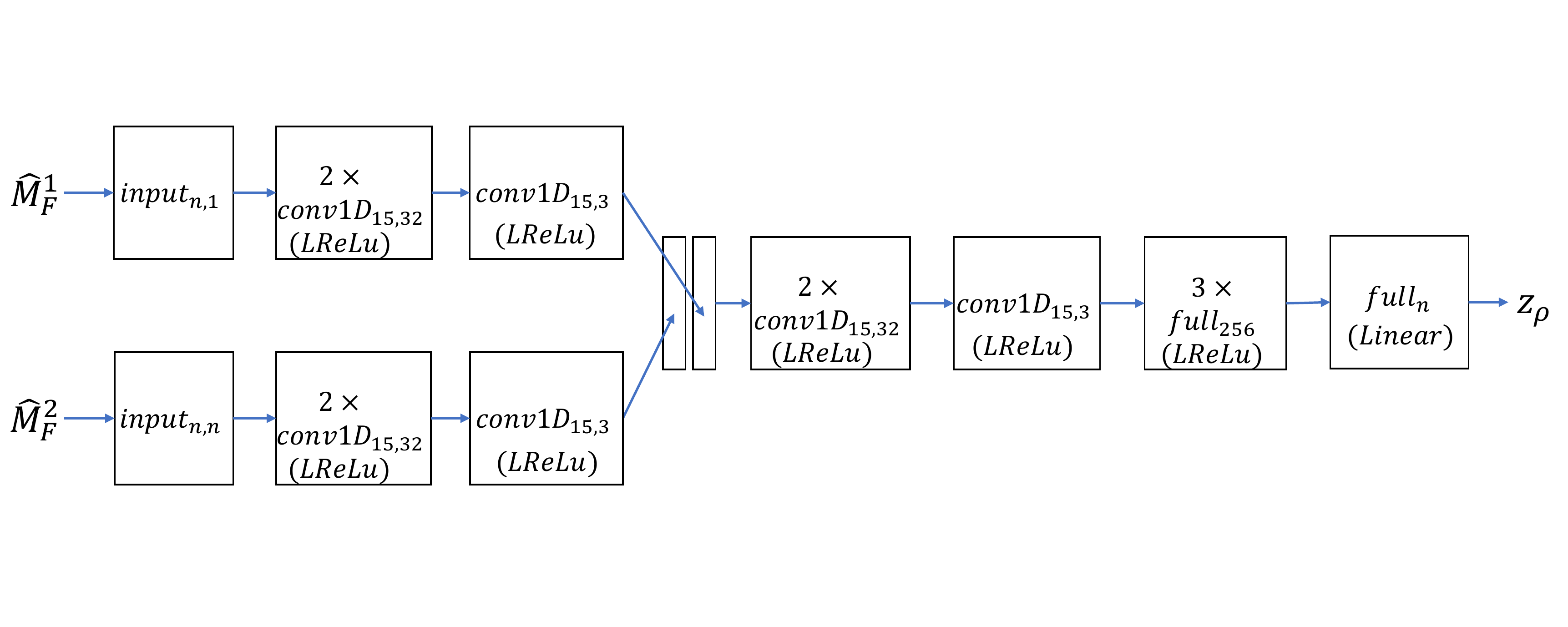}
    \caption{Architecture of $\xi^\rho_\theta$ in case of MRA}
    \label{fig:encoder_art}
\end{figure}

In the MRA case, we present the proposed architecture for the encoder $\xi_\theta$. An illustration of $\xi^\rho_\theta$ is presented in Figure~\eqref{fig:encoder_art}, and the same architecture is used for $\xi_\theta^v$. The input layers $\texttt{input1D}_{n,1}$ and $\texttt{input1D}_{n,n}$ take the moments as inputs. After a few layers of $\texttt{conv1D}$, we stack the output of the upper branch and lower branch in Figure~\eqref{fig:encoder_art} together into a 1D vector field of length $n$ and 6 channels. Then after a few more layers of CNN $\texttt{conv1D}$ and fully connected layers $\texttt{full}$, we output $z_\rho$.

\subsection{Cryo-EM}
The encoders $\xi^\rho_\theta$ and $\xi^v_\theta$ are very similar to the one presented in Figure~\eqref{fig:encoder_art} for MRA, except we replace all \texttt{conv1D} with \texttt{conv2D} with the same window sizes and channel numbers. As for $\hat v_\phi$, currently, it is chosen to be the FourierNet of~\cite{levy2022cryoai}. FourierNet finds success in representing the Fourier transforms of three-dimensional volumes of molecules and other volumes arising in nature, with values that often span multiple orders of magnitude. The main point of such a representation is that, instead of approximating $v(x)$ directly by an NN, it is often easier to approximate its Fourier coefficients $\widehat v(x)$ by an NN on $k$-space when $v(x)$ exhibits oscillatory patterns. This is also similar to the approach taken in \cite{cai2020phase} for solving high-frequency wave equations. More precisely, it lets
\begin{equation}
    \widehat v(k) \approx \widehat v_\phi(k) = a_{\phi_1}(k)\exp( ib_{\phi_2}(k))
\end{equation}
with two NNs $a_{\phi_1}(k)\in \mathbb{C}$ and $b_{\phi_2}(k)\in\mathbb{C}$ where $a_{\phi_1}$ gives the amplitude of the Fourier coefficients and $b_{\phi_2}$ gives the phase variations. By representing $v$ in Fourier domain instead of real domain, one can bypass the oscillatory pattern caused by the Fourier series $\exp(ikx)$ in $v(x) = \sum_k \widehat v(k) \exp(ikx)$. More details regarding the architecture, its effectiveness, and its memory requirements are provided in~\cite{levy2022cryoai}.

\end{document}